\documentclass[aps,pre,twocolumn,amssymb, amsmath,nofootinbib]{revtex4-1}
\usepackage{graphicx,latexsym,pifont,amsmath}
\usepackage{times} 
\usepackage{tikz}
\definecolor{labelkey}{cmyk}{.4,.2,0,0}
\usepackage{color}
\definecolor{Blue}{rgb}{0.00, 0.00, 1.00}
\definecolor{Red}{rgb}{1.00, 0.00, 0.00}

\newcommand{\nn}{\nonumber}

\newcommand{\Fig}[1]{\includegraphics[width=8.7cm]{./#1}}

\newlength{\bilderlength}

\def\be{\begin{equation}}
\def\ee{\end{equation}}

\def\bal{\begin{align}}
\def\eal{\end{align}}

\def\bea{\begin{eqnarray}}
\def\eea{\end{eqnarray}}

\usepackage{color}

\begin{document}

\title{Distribution of joint local and total size and of extension for avalanches in the Brownian force model}

\author{Mathieu Delorme, Pierre Le Doussal and Kay J\"org Wiese}
\address{CNRS-Laboratoire de Physique Th\'eorique de l'Ecole
Normale Sup\'erieure, 24 rue Lhomond, 75005 Paris, France
}

\begin{abstract}
\medskip
The Brownian force model (BFM) is a mean-field model for the local velocities during
avalanches in elastic interfaces of internal space dimension $d$, driven in a random medium.
It is exactly solvable via a non-linear differential equation. 
We study avalanches following a kick, i.e.\ a step in the driving force. We first recall the calculation of
the distributions of the global size (total swept area) and of the local jump size
for an arbitrary kick amplitude. We extend this calculation to the joint density 
of local and global sizes within a single avalanche,  in the limit of an infinitesimal kick. 
When the interface is driven by a single point we find new exponents $\tau_0=5/3$ 
and $\tau=7/4$, depending on whether the force or the displacement is imposed. 
We show that the extension of a ``single avalanche" along one internal direction (i.e.\ the total length
in $d=1$) is finite and we calculate its distribution, following either a
local or a global kick. In all cases it  exhibits a divergence $P(\ell) \sim \ell^{-3}$ at small $\ell$. 
Most of our results are tested in a numerical simulation 
in dimension $d=1$. 
\end{abstract}

\maketitle




\section{Introduction}

%
%
%
\label{s:Introduction}

In many physical systems the motion is not smooth, but proceeds by avalanches. This 
 jerky motion   is correlated over a broad range of space and time scales. 
Examples are  magnetic interfaces, fluid contact lines, crack fronts in fracture, 
strike-slip faults in geophysics
and many more \cite{ZapperiCizeauDurinStanley1998,LeDoussalWieseMoulinetRolley2009,DSFisher1998}.
These systems have been described using the model of an elastic interface slowly driven in a random medium.
This model is important for avalanches, both conceptually and in applications 
\cite{BonamySantucciPonson2008,PapanikolaouBohnSommerDurinZapperiSethna2011,DahmenSethna1996}. The full model of an interface of internal dimension $d$, in presence of realistic short-ranged disorder is  difficult to treat analytically, and requires methods such as the Functional Renormalization Group (FRG)
\cite{NattermannStepanowTangLeschhorn1992,NarayanDSFisher1993a,LeDoussalWiese2008c,LeDoussalWiese2011b,%
LeDoussalWiese2011a,DobrinevskiLeDoussalWiese2011b,LeDoussalWiese2012a,DobrinevskiLeDoussalWiese2014a}.

A simpler version of the model, the so-called Brownian force model (BFM) introduced in 
\cite{LeDoussalWiese2011b,LeDoussalWiese2011a,DobrinevskiLeDoussalWiese2011b,LeDoussalWiese2012a}
is very interesting in several respects. First it is exactly solvable, and several
avalanche observables have been calculated, as discussed below. 
Second, it was shown  
\cite{LeDoussalWiese2011a,LeDoussalWiese2012a}
to be the appropriate mean-field theory for the space-time statistics of the velocity field in a single avalanche for
$d$-dimensional 
interfaces close to the depinning transition for $d \geq d_{{\rm uc}}
$ with $d_{{\rm uc}} = 4$ for short ranged elasticity and $d_{{\rm uc}} = 2$ for long-ranged elasticity.
Remarkably, when considering the dynamics of the center of mass of the interface, it reproduces the results of
the simpler ABBM model, a toy model for a single degree of freedom (particle), introduced long ago on a phenomenological basis to describe Barkhausen  experiments (magnetic noise)\cite{AlessandroBeatriceBertottiMontorsi1990,AlessandroBeatriceBertottiMontorsi1990b} and much studied since \cite{ZapperiCizeauDurinStanley1998,Colaiori2008,DobrinevskiLeDoussalWiese2011b}. Last but not least, the BFM is the starting point for a calculation of avalanche observables beyond mean-field, using the FRG
in a systematic expansion in $d_{\rm uc}-d$ \cite{LeDoussalWiese2011a,LeDoussalWiese2012a}. 

The key property which makes the BFM (and the ABBM) model solvable is that 
the disorder is taken to be a Brownian random force landscape. Since it can be shown that
under monotonous forward driving the interface always moves forward (Middleton's theorem \cite{Middleton1992}), 
the resulting equation of motion for the velocity field is Markovian, and amenable to exact methods.

Despite being exactly solvable, the explicit calculation of avalanche observables in
the BFM requires solving a non-linear {\em instanton equation} and performing
Laplace inversions, which is not always an  easy task. Global avalanche properties, such as the probability distribution function (PDF) of
global size $S$, of duration, and of velocity have been obtained for arbitrary driving. Detailed
space time properties however are more difficult. In Ref.\ \cite{LeDoussalWiese2012a}
a finite wave-vector observable was calculated, demonstrating an asymetry in the
temporal shape. Although the distribution of local avalanche sizes $S_r$ has been 
obtained in some instances, this is not the case for the distribution of {\it the spatial extension $\ell$ of an avalanche}, 
i.e.\ the range of points which move during an avalanche,
an important observable  accessible in experiments. Note that even the fact that an avalanche has a finite extent, instead of an exponentially decaying tail in its spatial extension is a non-trivial result, which up to now was only proven for very large avalanches in the BFM  \cite{ThieryLeDoussalWiese2015}.

\begin{figure}[b]
\Fig{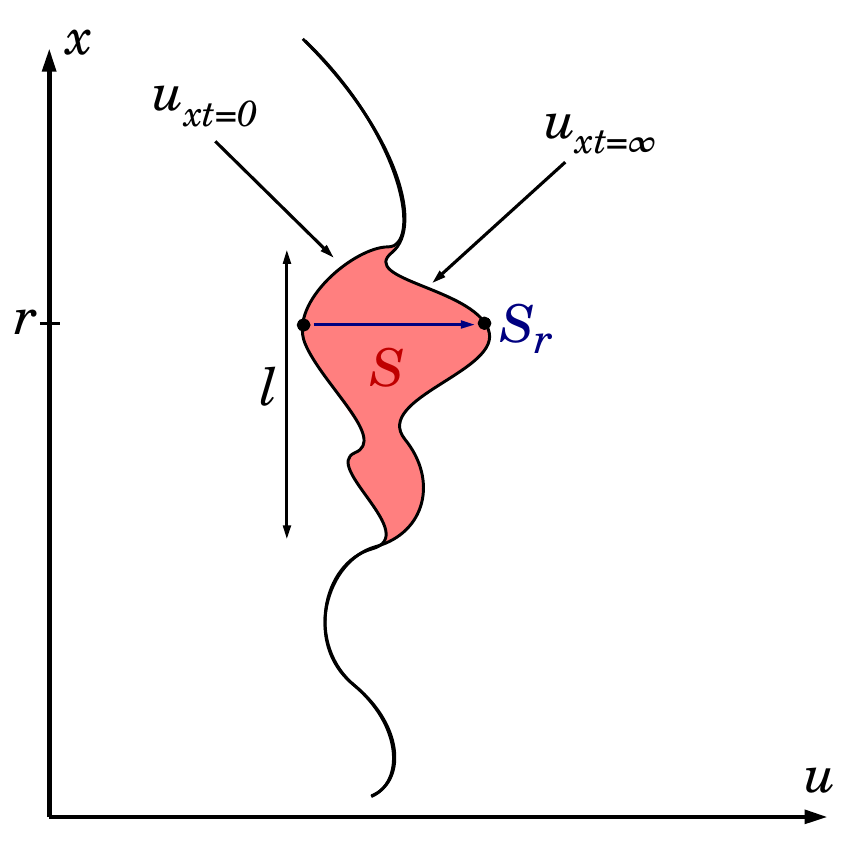}
\caption{An avalanche in $d=1$.}
 \label{Avalanche1d}
\end{figure}
The aim of this paper is to calculate further observables for the BFM which contain
information about local properties, such as the joint density of global and local avalanches, and 
the distribution of extensions. We consider various  protocols, where the interface is either driven uniformly in space
or   at a single point; in the latter case we identify new critical exponents.
We study avalanches following a kick, i.e.\ a step in the driving force.\\

 The article is structured as follows: 
In section \ref{sec:BFM} we recall the definition of the BFM and of
the main avalanche observables, together with the general method to 
obtain them from the instanton equation. Section \ref{sec:size}
  starts by recalling the calculation of
the distributions of the global size (total swept area) $S$ and of the local jump size $S_r$
of an avalanche, for an arbitrary kick amplitude. In Section \ref{subsec:joint} 
we extend this calculation to the joint density 
$\rho(S_r,S)$
of local and global size for single avalanches, i.e.\ in the limit of an infinitesimal kick. 
In Section \ref{sec:point} we study the case of an interface driven at
a single point. When the   {\it force} at this point is imposed, we 
find a new exponent $\tau_0=5/3$ for the PDF of the local jump $S_0$ at that
point. When the local {\it displacement} is imposed, we find a 
new exponent $\tau=7/4$ for the PDF of the global size $S$. 
In Section \ref{sec:extension} we show that the extension $\ell$ of a {\em single avalanche}
along one internal direction (i.e.\ the total length
in $d=1$) is finite; we calculate its distribution, following either a
local or a global kick. In all cases it   exhibits a divergence $P(\ell) \sim \ell^{-3}$ at small $\ell$, with the same prefactor. All these exponents can be found in Table~\ref{TableExponent}.
Finally, in Section \ref{sec:non-stat}  we study the {\em position} of the interface, which is a non-stationary process. We explain how the Larkin and BFM roughness exponents
emerge from the dynamics. 
Most of our results are tested in a numerical simulation of the equation of motion
in $d=1$.
\begin{table}\label{TableExponent}
\begin{tabular}{|c|c|c|}
	\hline
	Driving protocol & \hspace{0.4cm} Observable \hspace{0.4cm} & \hspace{0.2cm} Exponent \hspace{0.2cm} \\
	\hline \hline
	any force kick & global size $S$ & $\tau=3/2$ \\
	\hline
	uniform force kick & local size $S_0$ & $\tau_0=4/3$ \\
	\hline
	uniform force kick & $S_0$ at fixed $S$ & $\tau_0=2/3$ \\
	\hline
	localized force kick & local size $S_0$ & $\tau_0=5/3$ \\
	\hline
	local displacement imposed & global size $S$ & $\tau=7/4$ \\
	\hline
	any force kick & extension $\ell$ & $\kappa=3$ \\
		\hline
\end{tabular}
\caption{Summary of small-scale exponents for different distributions in the Brownian-Force Model, depending on the observable and the driving protocol.}
\end{table}

The technical parts of the calculations are presented in  Appendices \ref{app:airy} to \ref{app:nonstat}, together with general material about Airy, Weierstrass and Elliptic functions. A short presentation of the numerical methods is also included.

Finally note a complementary recent study of the BFM, where the joint PDF of the local avalanche size
at all points was obtained. From that,  the spatial shape
of an avalanche in the limit of large aspect ratio $S/\ell^4$ was derived \cite{ThieryLeDoussalWiese2015}. 

\section{Avalanche observables of the BFM}
\label{sec:BFM} 

\subsection{The Brownian Force Model}
In this paper, we study the Brownian Force Model (BFM) in space dimension $d$, defined as the stochastic differential equation (in the 
Ito sense)\,:
\be
\eta \partial_t \dot{u}_{xt}= \nabla_x^2 \dot{u}_{xt} 
+ \sqrt{2 \sigma \dot{u}_{xt}} \,\xi_{xt} + m^2(\dot{w}_{xt}- \dot{u}_{xt})\ .
\label{BFMdef}
\ee
This equation models the overdamped time evolution, with friction $\eta$, of the velocity field $\dot{u}_{xt} \geq 0$ of an interface with internal coordinate   $x \in \mathbb{R}^d$; the space-time dependence is denoted by indices $\dot{u}(x,t) \equiv \dot{u}_{xt}$. 
It is the sum of three contributions:
\begin{itemize}
\item short-ranged elastic interactions,
\item stochastic contributions from a disordered medium, where $\xi$ is a unit Gaussian white noise (both in $x$ and $t$)\,: 
\be
\overline{\xi_{xt}\xi_{x't'}} = \delta^d\!(x-x')\,\delta(t-t'),
\ee
\item a confining quadratic potential of curvature $m$, centered at $w_{xt}$, acting as a driving.
\end{itemize}
The driving velocity is chosen  positive, $\dot w_{xt} \geq 0$, a necessary condition for the model
to be well defined, as  it implies that $\dot{u}_{xt} \geq 0$ at all $t>0$ if $\dot{u}_{xt=0} \geq 0$. 

Equation \eqref{BFMdef}, taken here as a definition, can also be derived from the 
equation of motion of an elastic interface, parameterized by a position field (displacement field) $u_{xt}$ 
in a quenched random force field $F(u,x)$,
\be
\eta \partial_t u_{xt}= \nabla_x^2 u_{xt} + F\!\left(u_{xt},x\right) + m^2(w_{xt}- u_{xt})\ .
\label{BFMpos}
\ee
The random force field is a collection of independent one-sided Brownian motions in the $u$ direction
with correlator
\be
\overline{F(u,x)F(u',x')}=2 \sigma \delta^d(x-x')\min(u,u')\ .
\ee
Taking the temporal derivative $\partial_t$ of Eq.~\eqref{BFMpos}, and assuming forward motion of the interface,
one obtains Eq.~\eqref{BFMdef} for the velocity variable $\partial_t u_{xt} \equiv \dot{u}_{xt}$ 
(we use indifferently $\partial_t$ or a dot to denote time derivatives). 
The fact  that the equation for the velocity is  Markovian even for a quenched disorder is remarkable and results from
the properties of the Brownian motion. 

Details of the correspondence are given in \cite{DobrinevskiLeDoussalWiese2011b,LeDoussalWiese2012a} 
where subtle aspects of the position theory, and its links to the mean-field theory of 
realistic models of interfaces in short-ranged disorder via the Functional Renormalisation Group (FRG)
are  discussed. In the last section of this paper we will mention some properties of the position theory of
the Brownian force model. 

\subsection{Avalanches observables and scaling}

The BFM  \eqref{BFMdef} allows to study the statistics of avalanches as the 
dynamical response of the interface to a change in the driving. We consider solutions of \eqref{BFMdef} as a response to a driving of the form
\be \label{kick} 
\dot{w}_{xt} = \delta w_x ~ \delta(t)  \,\, ,  \,\, \delta w_x \geq 0  \,\,,  \,\,\delta w = L^{-d} \int_x \delta w_x >0 \ .
\ee
The initial condition is
\be \label{init}
\dot{u}_{xt=0} =0\ .
\ee 
This solution describes an 
avalanche which starts at time $t=0$ and ends when $\dot u_{xt}=0$ for all $x$. The time at which the avalanche ends, also called avalanche duration, was studied in \cite{DobrinevskiPhD}  and its distribution  given in various situations.

Within the description \eqref{BFMpos}, i.e.\ in the position theory, it corresponds to an interface pinned, i.e.\ at rest in a metastable state at $t<0$, it is submitted at $t=0$ to a jump in the total applied force $m^2 \delta w$. More
precisely, the center of the confining potential jumps at $t=0$ from $w_{x}$ (where it was for $t<0$) to $w_{xt=0^+} = w_{x} + \delta w_x$ (where it stays for all $t>0$). As a consequence, the  interface  moves forward (since $\delta w_x \geq 0$) up to a new metastable state. This is represented in figure \ref{Avalanche1d}, where $u_{xt=0}$ is the initial metastable state and $u_{xt=\infty}$ is the new metastable state at the end of the avalanche. In fact,   as we will see from the distribution of   avalanche durations, the new metastable state is reached almost surely in a finite time.
For   details on these metastable 
states and the system's preparation see \cite{DobrinevskiLeDoussalWiese2011b,LeDoussalWiese2012a}. \\

\begin{figure}[t]
\Fig{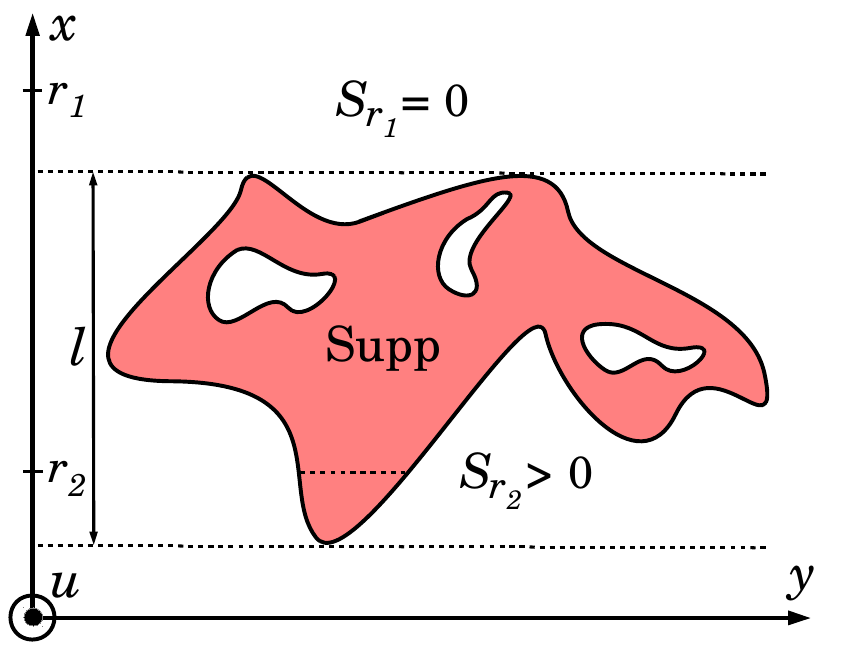}
\caption{An avalanche in $d=2$; the transverse direction is orthogonal to the plane of the figure and the colored zone corresponds to the support of the avalanche.}\label{Avalanche2d} 
\end{figure}

We   now discuss the   avalanche observables  at the center of this paper. They can be computed from the solution of \eqref{BFMdef} given \eqref{kick} and \eqref{init}; they are represented in figure \ref{Avalanche1d} for a more visual definition in the case $d=1$.
\begin{itemize}
\item Global size of the avalanche:
\be
S=\int_{x\in \mathbb{R}^d} \int_{0}^{\infty} \dot{u}_{tx} \, dt\ .
\label{GlobalSize}
\ee
This is the total area swept by the interface during the avalanche.
\item Local size of the avalanche:
\be
S_r = m^{-1}\int_{x \in \{r\}\times \mathbb{R}^{d-1}} \int_{0}^{\infty}  \dot{u}_{tx} \, dt\ .
\label{LocalSize}
\ee
This is the size of the avalanche localized on a hyperplane, where one of the internal coordinates is $r$; the factor $m^{-1}$ allows to express $S$ and $S_r$ using the same units (see below). 
In $d=1$ this yields $S_r= m^{-1} \int_{0}^{\infty} \dot{u}_{tr}\, dt$, i.e.\ the transversal jump at the point $r$ of the interface. For $d>1$ the variable $r$ is still   one-dimensional,   and $S_r$ the total displacement in a hyperplane of the interface.

\item Avalanche extension:

For $d=1$, the extension (denoted $\ell$) of an avalanche is the lenght of the part of the interface which (strictly) moves during the avalanche. The generalisation to avalanches of a $d$-dimensional interface is done with the definition
\be
\ell=  \int_{-\infty}^{\infty} dr~ \theta(S_r > 0)\ ,
\label{extension}
\ee
where $\theta$ is the Heaviside function. Note that even for a $d$-dimensional interface, the extension $\ell$ is   a unidimensional observable (\textit{cf.} figure \ref{Avalanche2d}).
\end{itemize}
Note that
\be
S_r >0\  \Leftrightarrow \ {\rm Supp} \bigcap \{r\}\times \mathbb{R}^{d-1} \neq \emptyset
\ee
where ${\rm Supp}$ denotes all the points of the interface moving during an avalanche (i.e.\ its support). 

We   use natural scales (or units) to switch to dimensionless expressions, both for the (local and global) avalanche size  $S_m$,  as for the time $\tau_m$ expressed as
\be
S_m = \frac{\sigma}{ m^4} \ , \qquad  \tau_m = \frac{\eta}{m^2}\ .
\label{units}
\ee
The extension,   a length in the internal direction of the interface, is  expressed in units of $m^{-1}$. This is equivalent to setting $m=\sigma=\eta = 1$. All expressions below, except explicit mention, are expressed in these units.

While $S_m$ is the large-size cutoff for avalanches, there is generically also a small-scale cutoff. As in the BFM the disorder is scale-invariant (by contrast with more realistic models with short-ranged smooth disorder),   it is the
 increment in the driving $\delta w$ which sets the small-scale cutoff for the local and global size of avalanches.
They scale   as $\min(S) \sim \delta w^2$ (global size) and $\min(S_r) \sim \delta w^3$ (local size).

\vspace{0.2cm}
{\it Massless limit}
\vspace{0.2cm}

There are cases of interest where the mass $m \to 0$. This can be
defined from the equations of motion (\ref{BFMdef}) and (\ref{BFMpos}) with the changes
\be
\begin{split}
m^2(w_{xt}- u_{xt}) &\to f_{xt} \\
m^2(\dot{w}_{xt}- \dot{u}_{xt}) &\to \dot f_{xt}\ .
\end{split}
\ee
In that case it is natural to consider driving with a given force $f_{xt}$, rather 
than by a parabola. The definition of the observables is the same except that
the factor of $m^{-1}$ is not added in the definition (\ref{LocalSize}). 
To bring $\sigma$ and $\eta$ to unity, we  then
define  time in units of $\eta$ and displacements 
$u$ in units of $\sigma$. The results will still have an unfixed 
dimension of length. In some of them,  the system size $L$
  leads to dimensionless quantities (it also acts as
a cutoff for large sizes, although we will not use this explicitly). 

\subsection{Generating functions and instanton equation}

To compute the distribution of the observables presented above, we   use a result from \cite{DobrinevskiLeDoussalWiese2011b,LeDoussalWiese2012a} which allows us to express the average over the disorder of generating functions (Laplace transforms) of $\dot{u}_{xt}$, solution of \eqref{BFMdef}. In dimensionless units, this result reads
\be
G[\lambda_{xt}]=\left\langle \exp \left(\int_{xt} \lambda_{xt}\dot{u}_{xt} \right) \right\rangle = e^{ \int_{xt} \dot{w}_{xt} \tilde{u}_{xt} }\ .
\label{generating}
\ee
Here $\langle\cdots \rangle$ denotes the average over disorder. $\tilde{u}$ is a solution of the differential equation (called instanton equation)
\be
\partial_{x}^2\tilde{u} +  \partial_{t}\tilde{u}-\tilde{u}+ \tilde{u}^2 =- \lambda_{xt}\ .
\label{instanton2}
\ee
Since avalanche observables that we consider are  integrals of the velocity field over all times (c.f.\ observable definitions above), the sources $\lambda_{xt}$ we need in \eqref{generating} are {\em time independent}. Thus we only need to solve the space dependent, but time independent, instanton equation
\be
\tilde{u}_x''-\tilde{u}_x+ \tilde{u}^2_x =- \lambda_{x}\ .
\label{instanton}
\ee
The prime denotes derivative w.r.t $x$. In the massless case discussed above,  the term $- \tilde{u}_x$ is absent, all other terms are identical. 

The global avalanche size implies   a uniform source in the instanton equation: $\lambda_x = \lambda$, while  the local size implies a localized source $\lambda_x =  \lambda \delta^1(x)$. To obtain information on the extension of avalanches, we need to consider a source localized at two different  points in space,  $\lambda_x = \lambda_1 \delta(x-r_1) + \lambda_2 \delta(x-r_2)$.

This instanton   approach, which derives from the Martin-Siggia-Rose formulation of \eqref{BFMdef}, allows us to compute exactly disorder averaged    observables for any form of driving, by solving a ``simple" ordinary differential equation, which depends on the observable we want to compute, i.e.\ on $\lambda_{xt}$, but not on the form of the driving $\dot{w}_{xt}$. For a   derivation of \eqref{generating} and \eqref{instanton2} we refer to \cite{DobrinevskiLeDoussalWiese2011b}.


\section{Distribution of avalanche size}
\label{sec:size} 

\subsection{Global size}

As defined in \eqref{GlobalSize} the global size of an avalanche is the total area swept by the interface. Its PDF was 
calculated in \cite{DobrinevskiLeDoussalWiese2011b,LeDoussalWiese2011a,LeDoussalWiese2012a} and reads, in
dimensionless units, 
\be
P_{\delta w} (S) = \frac{\delta \hat{w}}{2 \sqrt{\pi}S^{\frac{3}{2}}}e^{-\frac{(S-\delta \hat{w})^2}{4S}}\ .
\label{GlobalDistrib}
\ee
Here $\delta \hat{w} = L^{d} \delta w$. This result does not depend on the spatial form of the driving (it can be localized, uniform, or anything in between), as long as it is applied as a force on the interface. Driving by imposing a specific displacement at one point of the interface is another interesting case that leads to a different behavior, see Section \ref{sec:imposed}. 

\begin{figure}[t]
\Fig{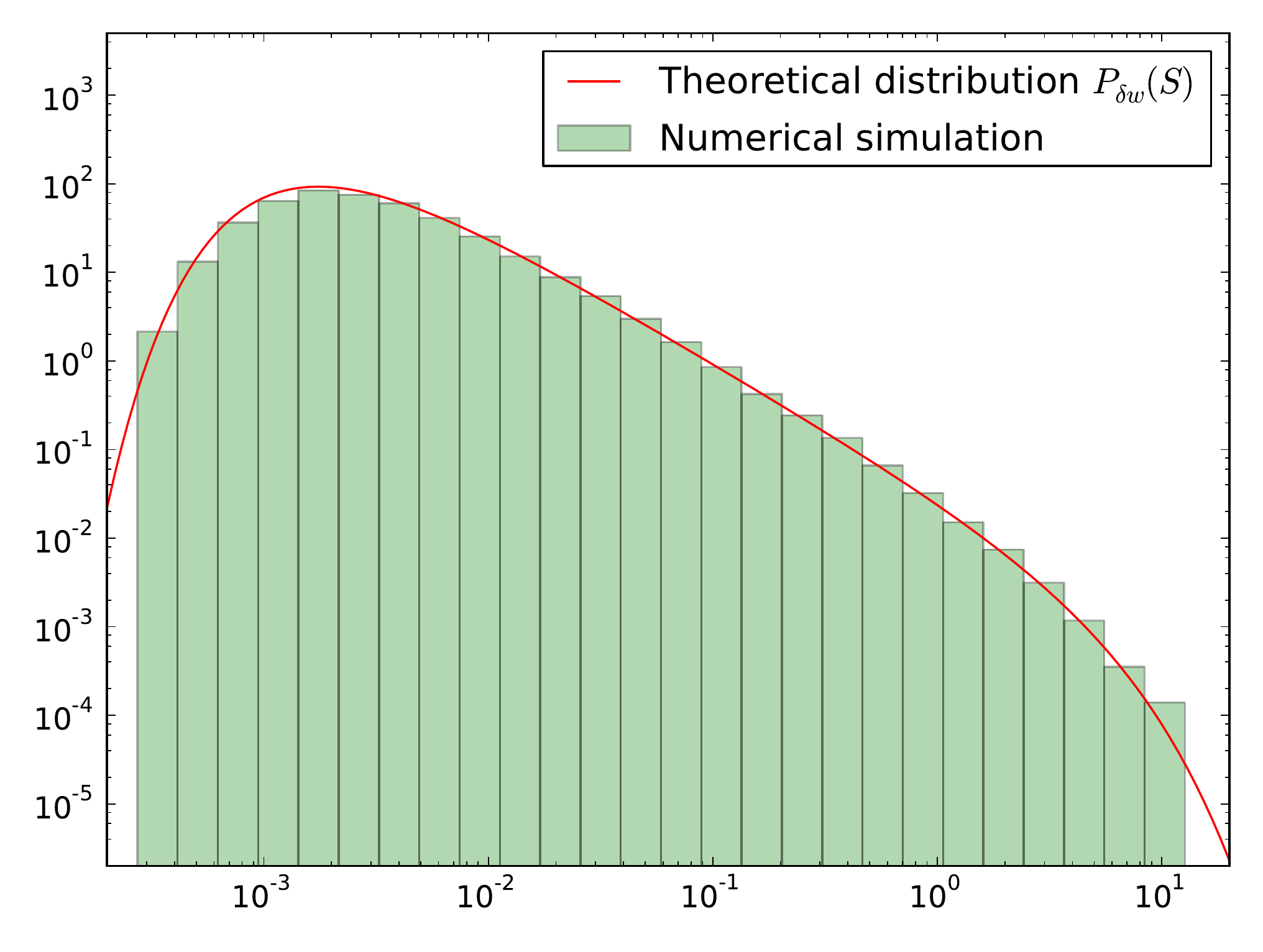}
\caption{Green histogram : global avalanche-size distribution from a direct numerical simulation of a discretized version of Eq.~\eqref{BFMdef} with parameters : $N=1024, m=0.01, df = m^2 \delta w = 1$ and $dt=0.05$. Red line : theoretical result given in Eq.~\eqref{GlobalDistrib}. For details about the simulation see appendix \ref{a:Numerics}.}\label{GlobalSizeFig} 
\end{figure}

We can test this against a direct numerical simulation of the equation of motion \eqref{BFMdef}. There is   excellent agreement over 5 decades, with no fitting parameter, see Fig.~\ref{GlobalSizeFig}.

Avalanches have the property of infinite divisiblity,  i.e.\ they are a Levy process. This can be written as an equality in distribution, i.e.\ for probabilities, 
\be
P_{\delta w_1} * P_{\delta w_2} \overset{d}{=}P_{\delta w_1+ \delta w_2}\ .
\ee
It implies that we can extract from the probability distribution \eqref{GlobalDistrib} the {\em single avalanche} density per unit $\delta w$ that we denote $\rho(S)$ and which is defined as
\be
P_{\delta w} (S) \underset{\delta \hat w \ll 1}{\simeq} \delta w \,\rho(S)\ .
\ee
This avalanche density contains the same information as the full distribution \eqref{GlobalDistrib}; its expression is
\be
\rho(S) = \frac{L^d}{2 \sqrt{\pi} S^{\frac{3}{2}}}e^{-\frac{S}{4}} \sim S^{-\tau}\ .
\label{GlobalDensity}
\ee
It is proportional to the system volume since avalanches occur anywhere along
the interface. It defines the avalanche exponent $\tau=\frac{3}{2}$ for the
BFM. Due to  the divergence when $S \rightarrow 0$ it is not
normalizable (it is not a PDF), but as the interface follows on average the confining parabola, it has the following property
\be
\int_{0}^{\infty} \!\!\!dS\,S \rho(S)  = L^d \ .
\ee
In this picture, typical, i.e.\ almost all avalanches are of  vanishing size,  $S\approx 0$, or more precisely
 $S\le \delta \hat{w} ^2$, but moments of
avalanches are dominated by   non-typical large avalanches
(of order $S_m$). 

\subsection{Local size}

We now investigate the distribution of local size $S_r$ as defined in Eq.~\eqref{LocalSize}. We have to specify   the form of the kick;  we   start with one uniform (in $x$): $\delta w_x = \delta w$ for all $x \in \mathbb{R}$. In this case the system is translationnaly invariant, and we can choose  $r=0$, as any local size will have the same distribution.

The distribution of $S_0$ is obtained by solving Eq.~$\eqref{instanton}$ with the source $\lambda_x =\lambda \delta(x)$, and then computing the inverse Laplace transform with respect to $\lambda$ of $G(\lambda) = \exp (\delta w \int_x \tilde{u}^{\lambda} )$, where $\tilde{u}^{\lambda}$ is the instanton solution (depending on $\lambda$). This has been done in \cite{LeDoussalWiese2012a}; the final result is
\be \label{LocalDistrib}
\begin{split}
\!\!P_{\delta w}(S_0)&= \frac{2\times 3^{\frac{1}{3}}}{S_0^{\frac{4}{3}}}e^{6 \delta \hat{w}} \delta \hat{w}\, \text{Ai}\!\left( \left( \frac{3}{S_0}\right)^{\!\frac{1}{3}}(S_0+2 \delta \hat{w})\right)\\
&\simeq_{\delta \hat w \ll 1} \delta w \frac{2 L^{d-1}}{\pi S_0} \text{K}_{\frac{1}{3}}\!\left(\frac{2 S_0}{\sqrt{3}} \right)\ .
\end{split}
\ee
Here $\delta \hat{w}= L^{d-1} \delta w$, Ai is the Airy function, and K the Bessel function. 
We use that $\text{Ai}(x)=\frac{1}{\pi} \sqrt{\frac{x}{3}} K_{1/3}(\frac{2}{3} x^{3/2})$ for $x>0$. This distribution has again the property of infinite divisibility, which is far from obvious on the final results but, can  be checked numerically.

The small-$\delta w$ limit defines the density per unit $\delta w$ of the local sizes of a ``single avalanche", which is given by
\be\label{LocalDensity}
\begin{split}
 \rho(S_0) & = \frac{2 L^{d-1}}{\pi S_0} \,\text{K}_{\frac{1}{3}}\! \!\left(\frac{2 S_0}{\sqrt{3}} \right) \\
& \simeq_{S_0 \ll 1} L^{d-1} \frac{\sqrt[6]{3}\,\Gamma
   (1/3)}{\pi  {S_0}^{4/3}} \sim S_0^{- \tau_\phi} \ .
\end{split}
\ee
Its small-size behavior defines the local size exponent $\tau_\phi=\frac{4}{3}$ for the BFM. 

\begin{figure}[t]\Fig{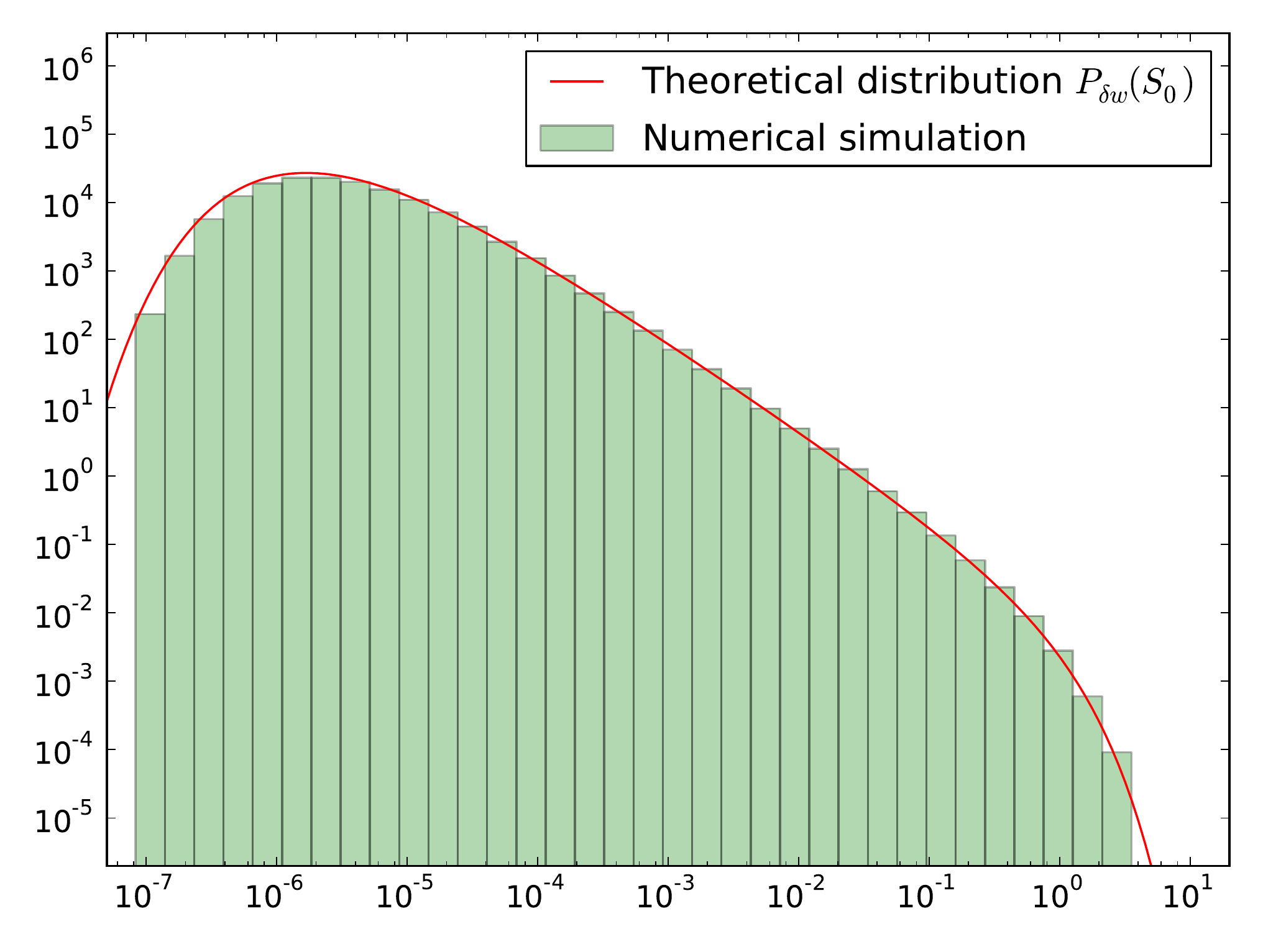}
\caption{Green histogramm: Local avalanche-size distribution from a direct numerical simulation of a discretized version of \eqref{BFMdef}with parameters  $N=1024, m=0.01, df = m^2 \delta w = 1$, and $dt=0.05$. Red line: the theoretical result given in Eq.~\eqref{LocalDistrib}. For details about the simulation see appendix \ref{a:Numerics}.}\label{LocalSizeFig} 
\end{figure}

The distribution \eqref{LocalDistrib}, or the density \eqref{LocalDensity}, can be compared to the results of direct numerical simulations of the BFM, and the agreement is very good over 7 decades, {\em without any fitting parameter}, c.f. Fig.~\ref{LocalSizeFig}.

Another interesting property is that the tail of large local sizes behaves as $\rho(S_0) \simeq_{S_0 \gg 1} S_0^{-3/2} e^{-2S_0/\sqrt{3}}$, i.e.\ with the same power-law exponent in the pre-exponential factor as the global size.

\subsection{Joint global and local size}
\label{subsec:joint} 

We now extend these results with a new calculation of the joint density of local   and global sizes. 
Consider $P_{\delta w}(S_0,S)$, the joint PDF of local size $S_0$ and global size $S$,  following a uniform kick
$\delta w$. For arbitrary $\delta w$ it does not admit a simple explicit form (see  
Appendix \ref{sec:joint}). We thus again
consider the ``single avalanche" limit $\delta w \rightarrow 0$. It defines the joint density $\rho(S,S_0)$, via
$P_{\delta w}(S_0,S) \simeq \delta w\, \rho(S_0,S)$, which we now calculate. Equivalently one can
consider the conditional probability $P_{\delta w}(S_0|S)$ of the local size, given that the global size is $S$. In the
limit $\delta w \to 0$ these two objects are   related by
\be
P_{0^+}\!(S_0|S)= \frac{\rho(S_0,S)}{\rho(S)} \ ,
\ee
where $\rho(S)$ is given in Eq.\ (\ref{GlobalDensity}); the two factors of $\delta w$ cancel. 
For simplicity we   discuss the result for 
$P_{0^+}\!(S_0|S)$. While both $\rho(S)$ and $\rho(S_0,S)$ are not probabilities, i.e.\ they cannot be normalized to one, we will show that the conditional probability $P_{0^+}\!(S_0|S)$ is   well-defined, and   normalized to unity.
 
A natural decomposition of this conditional PDF is
\be \label{dec} 
P_{0^+}\!(S_0|S) = \hat P_{0^+}\!(S_0|S) + \delta(S_0) \left(1  -  \int_{u>0}\!\!\!\hat P_{0^+}\!(u|S) \right)\ .
\ee
The first term is the smooth part defined for $S_0>0$ which comes from
the avalanches containing the point $ r=0$. The second term arises from all avalanches
which do not contain the point $ r=0$. This term contains a substraction so that the
total probability  is normalized to unity, $\int_{S_0} P_{0^+}(S_0|S) =1$, as it should be.

The smooth part is calculated   using the instanton-equation approach. 
The details are given Appendix \ref{sec:joint}.
The final result takes the scaling form
\be\label{jointdensity}
\hat P_{0^+}\!(S_0|S) = \frac{1}{L} {4 \times 3^{2 \over 3} \over S_0^{2 \over 3}} e^{-\frac{2}{3}\alpha^3 } \Big[\alpha\, \text{Ai} \big( \alpha^2 \big)  -  \text{Ai}'\big(\alpha^2\big) \Big]
\ee
with
\be\label{alphaDef}
\alpha := {3^{2 \over 3} S_0^{4 \over 3} \over S}\ .
\ee
The factor $1/L$ is natural since only a fraction of order $  1/L$ of 
avalanches contains the point ${r=0}$. As written, this smooth part is not normalized.
Its integral is equal to the probability $p$ that the point $S_0$ has moved
(i.e.\ $S_0>0$) during an avalanche, for which we find
\be 
p := \int_0^{\infty}\!\!\!dS_0\, \hat P_{ 0^+}(S_0|S) = \frac{S^{\frac{1}{4}}}{L} \frac{3 \Gamma\!\left( { 1 \over 4} \right) }{\sqrt{\pi}}\ .
\ee  
The scaling of this probability with size shows that in a single avalanche only a finite
portion of the interface is moving. If we assume statistical translational invariance
we deduce that
\be
p = \langle \ell \rangle_S /L\ ,
\ee
where $\ell$ is the extension defined in (\ref{extension}), and $\langle \ell \rangle_S$ its mean
value conditioned to the global size $S$. Hence we deduce that
\be
\langle \ell \rangle_S =  \frac{3 \Gamma\!\left( { 1 \over 4} \right) }{\sqrt{\pi}}  S^{\frac{1}{4}}\ .
\ee
In the following sections we will in fact calculate the 
PDF of the extension $\ell$.
\begin{figure}[t]
\Fig{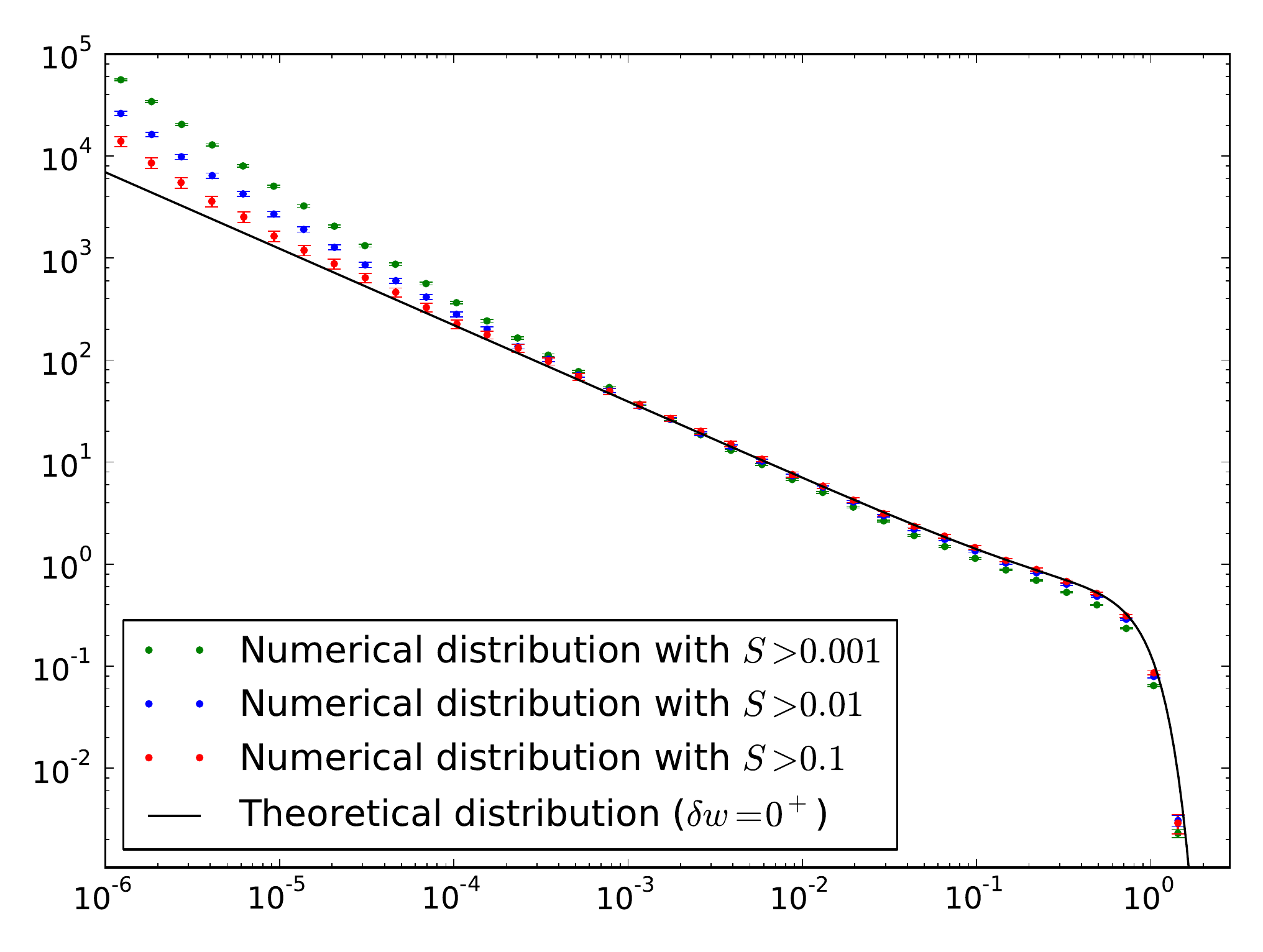}
\caption{Distribution of $\alpha$, defined in Eq.~\eqref{alphaDef}, from numerical simulations ($N=1024,m=0.02,\delta w=10, dt=0.01$). This is  compared to the theoretical prediction \eqref{alphaDistrib}. Keeping only   large-size avalanches,  this converges  (without any adjustable parameter) to the $\delta w =0^+$ result. }\label{AlphaDistribFig} 
\end{figure}

By dividing by $p$, we can now define a genuine normalized PDF for $S_0$, $\tilde P_{0^+}\!(S_0|S)$, conditioned to both $S$ and $S_0>0$, so that the decomposition (\ref{dec}) becomes
\be
P_{0^+}\!(S_0|S) = p\,\tilde P_{0^+}\!(S_0|S) + \delta(S_0) (1-p)  \ .
\ee
Explicitly
\be\label{S0DistribSfixed}
\tilde P_{0^+}\!(S_0|S) = \frac{4 \sqrt{\pi} e^{-\frac{2}{3}\alpha^3 } }{3^{1 \over 3} \Gamma\!\left(\frac{1}{4}\right) S_0^{2 \over 3} S^{\frac{1}{4} } }  \Big[\alpha\, \text{Ai} \big( \alpha^2 \big)  -  \text{Ai}'\big(\alpha^2\big) \Big]\ ,
\ee
with $\alpha$  defined in Eq.~\eqref{alphaDef}. 
It is 
now normalized to unity, $\int_{S_0>0} \tilde P_{0^+}(S_0|S) =1$. 
One sees that the typical local size scales as $S_0 \sim S^{3/4}$.
Computing the first moment we find its conditional average 
to be $\langle S_0 \rangle_{S,S_0>0} = \frac{\sqrt{\pi }}{3 \Gamma \left(1/4\right)} S^{3/4}$.
Its PDF has two limiting behaviors,
\be
\tilde P_{0^+}\!(S_0|S) \simeq \left\{
\begin{array}{ll}
 \dfrac{e^{-\frac{12 S_0^4}{S^3}}}{ \Gamma(\frac{5}{4}) S^{\frac{3}{4}}} &\text{ for } S_0 \gg S^{\frac{3}{4}}\\
\dfrac{ \sqrt{\pi}} { 3^{\frac{2}{3}} \Gamma(\frac{1}{3}) \Gamma(\frac{5}{4}) 
S_0^{\frac{2}{3}}  S^{\frac{1}{4}} }&\text{ for }S_0 \ll S^{\frac{3}{4}}\ .
\end{array}\right.
\ee
The first one shows that the probability of avalanches which are   ``peaked" at $r=0$ 
decays very fast. The second shows an integrable divergence at small $S_0$
with an exponent $2/3$. Comparing, for instance, with the
behavior of the local size density (\ref{LocalDensity}), we see that conditioning
on $S$ yields a rather different behavior and exponent.

It is   interesting to note that changing variables  in Eq.~\eqref{S0DistribSfixed} from $S_0$ to $\alpha$, defined in \eqref{alphaDef}, gives
\be \label{alphaDistrib}
\tilde P_{0^+}\!(\alpha|S) = \frac{\sqrt{3 \pi} e^{-\frac{2}{3}\alpha^3 } }{  \Gamma\!\left(\frac{1}{4}\right) \alpha^{\frac{3}{4}} }  \Big[\alpha\, \text{Ai} \big( \alpha^2 \big)  -  \text{Ai}'\big(\alpha^2\big) \Big]
\ ,
\ee
which is now independant of $S$, and thus easier to test numerically as it does not require any conditionning. Figure \ref{AlphaDistribFig} shows the agreement of these predictions with numerical simulations, in the limit of large $S$ which is equivalent to $\delta w=0^+$ as used in the theoretical derivation.


\subsection{Scaling exponents}

Let us now discuss the various exponents obtained until now. They are consistent with the
usual scaling arguments for interfaces. If an avalanche has an extension of order $\ell$ 
(in the codirection of the hyperplane over which the local size is calculated), the
transverse displacement scales as $u \sim \ell^\zeta$. Here the roughness exponent $\zeta$
for the BFM with SR elasticity is
\be
\zeta_{{\rm BFM}} = 4 -d \ .
\ee
The  avalanche exponent for the global size follows the Narayan-Fisher (NF)  prediction \cite{NarayanDSFisher1993a}
\be\label{NF}
\tau = 2 - \frac2{d + \zeta} ~\stackrel{\rm BFM}{-\!\!\!\longrightarrow}~ \frac{3}{2}  \ .
\ee
The global size then scales as $S \sim \ell^{d + \zeta}$, since  
all $d$ internal directions are equivalent, and the transverse response scales with the roughness exponent $\zeta$.
In turn this gives $\ell \sim S^{1 \over d+ \zeta}$. In the BFM with SR elasticity
this   leads to $\ell \sim S^{1/4}$ as found above.

Similarly, the local size, defined here   as the avalanche size inside a $d_{\phi}$-dimensionel subspace, is $S_0 \sim \ell^{d_\phi + \zeta}$ , leading to
a generalized NF value $\tau_\phi = 2 - \frac{2}{d_\phi + \zeta}$. 
In the BFM we have focused on the case $d_\phi=d-1$ (i.e. the subspace is an hyperplane), hence $d_\phi+\zeta=3$ and 
the local size exponent becomes $\tau_\phi = 4/3$.
It also implies $S_0 \sim \ell^3$, hence $S_0 \sim S^{3/4}$ as found above. 

\section{Driving at a point: avalanche sizes}

\label{sec:point} 

Here we briefly study  avalanche sizes for an interface driven only in a small
region of space, e.g.\ at a point. There are two main cases:

\begin{itemize}

\item the local force on the point is imposed, which in our framework means to
consider a local kick $\delta w_x= \delta w \,\delta(x)$. In the massless setting 
it amounts to use $f_x = \delta f\, \delta(x)$,

\item the displacement $u_{x=0,t}$ of one point of the interface is imposed.

\end{itemize}

As we now see this leads to different universality classes and exponents. 

\subsection{Imposed local force}

Consider an avalanche following a local kick at $x=0$, i.e.\ $\delta w_x=\delta w_0 \delta(x)$. 

In the BFM the distribution of the {\it global size} of an avalanche 
does not depend on whether the kick is local in space or not. One still obtains \cite{LeDoussalWiese2012a} the
global-size distribution as given in Eq.~(\ref{GlobalDistrib}) with $\delta \hat w = \int_x \delta w_x=\delta w_0$.

The distribution of the {\it local size at the point of the kick 
} is more interesting. The 
calculation is performed in   Appendix \ref{app:local}. For simplicity
we restrict   to $d=1$, the general case can be obtained as above by
inserting factors of $L^{d-1}$. The full result for the PDF, $P_{\delta w_0}\!(S_0)$, 
is given in (\ref{resloc})
and is bulky. In the limit $\delta w_0 \to 0$ it simplifies. Noting $P_{\delta w_0}\!(S_0)
\simeq \delta w_0 \rho(S_0)$,  the 
corresponding local-size density becomes
\be
\rho(S_0) = -  \frac{1}{3^{1/3} S_0^{5/3}} \text{Ai}'\!\left(3^{1/3} S_0^{2/3}\right)\ .
\ee
At small $S_0$, or equivalently in the massless limit at fixed $\delta f_0 = m^2 \delta w_0$, it diverges as
\be
 \rho(S_0)  \underset{S_0\ll1}{\simeq} \frac{S_0^{- 5/3}}{3^{2/3} \Gamma(1/3)} \sim S_0^{-\tau_{0,{\rm loc. driv.}} }\ .
\ee
This leads to a {\it new avalanche exponent} 
\be
\tau_{0,{\rm loc. driv.}}=\frac{5}{3}\ .
\ee
The cutoff at   small size is given by the driving,  $S_0 \sim \delta w_0^{3/2}$. At large $S_0$ the PDF is cut by the scale $S_m\equiv 1$ and decays as
\be
 \rho(S_0) \underset{S_0\gg1}{\simeq}\frac{S_0^{-3/2}}{2 \sqrt{\pi} 3^{1/4}} e^{-2 S_0/\sqrt{3}} \ .
\ee

\subsection{Imposed displacement at a point}
\label{sec:imposed} 

We   analyze the problem in the massless case. To impose the 
displacement at point $x=0$ we replace in the equation of motion
(\ref{BFMdef}) and (\ref{BFMpos}), $m^2 \to m^2 \delta(x)$.
Hence there is no global mass, but a local one to drive the
interface at a point. To impose the displacement, we consider the limit $m^2 \to \infty$.
In that limit   $u_{x=0,t}=w_{0,t}$, and the local size of the
avalanche $S_0$ is equal to $\delta w_0$.

While the local size $S_0$ is fixed by the driving, we can
calculate the distribution of global sizes. It is obtained in
  Appendix \ref{imposeddispl} using an instanton 
equation with a Dirac mass term. It can be   mapped
onto the same instanton equation as studied for the
joint PDF of local and global sizes. The Laplace-transform of the result for the
PDF  is given  in Eq.~(\ref{LT1}). Its small-driving limit,
i.e.\ the density, is
\be
\rho(S) = \frac{\sqrt{3}}{\Gamma(1/4) S^{7/4}}  \sim S^{-\tau_{\rm loc. driv.}} 
\ee 
with a distinct exponent 
\be
\tau_{\rm loc. driv.} = \frac{7}{4} \ .
\ee

\section{Distribution of avalanche extensions}
\label{sec:extension} 

In this section we study the distribution of avalanche extensions. 
In the BFM they can be calculated analytically. We start by recalling standard scaling arguments. 

\subsection{Scaling arguments for the distribution of extensions}

As mentioned in the last section, we expect that 
the global size $S$ and the extension $\ell$ of avalanches are related by the scaling relation
\be
S \sim \ell^{d + \zeta}
\ee
in the region of small avalanches $S \ll S_m$ (in dimensionfull units). 
From the definition of the avalanche-size exponent 
\be
P(S) \sim S^{-\tau}
\ee 
and using the change of variables $P(S) dS = P(\ell) d\ell$
we find
\be 
P(\ell) \sim \ell^{- \kappa} \, \text{ with } \, \kappa = 1 + (\tau-1) (d+\zeta) \ .
\ee  
Using the value for $\tau$ from the NF relation (\ref{NF}) we obtain
\be 
\tau = 2 - \frac{2}{d+\zeta}\ .
\ee  
For SR elasticity, this yields
\be
 \kappa = d + \zeta -1\ .
\ee
The prediction   for the BFM   is that $\zeta_{\rm BFM}=4-d$ and $\tau_{\rm BFM}=3/2$,
which leads to 
\be
\kappa_{\rm BFM}=3
\ee
in all dimensions. We will now check this prediction from the scaling relations 
with exact calculations on the BFM model in $d=1$. 

\subsection{Instanton equation for two local sizes}

If we want to investigate the joint distribution of two local sizes at points $r_1$ and $r_2$, 
we need to solve the instanton equation with two local sources,
\begin{equation}
\tilde{u}''_x -\tilde{u}_x+ \tilde{u}^2_x =- \lambda_1 \delta(x-r_1) - \lambda_2 \delta(x-r_2)\ .
\label{InstantonTwoSources}
\end{equation}
This solution is difficult to obtain for general values of $\lambda_1$ and $\lambda_2$.
Nevertheless $\lambda_{1,2} \rightarrow - \infty$ is an interesting solvable limit,  and sufficient to compute the extension distribution. Let us denote by $\tilde{u}_{r_1,r_2}(x)$ a solution of Eq.~\eqref{InstantonTwoSources} 
with $r_1<r_2$ in this limit $\lambda_{1,2} \rightarrow - \infty$. It allows to express the probability that two local sizes in an avalanche following an arbitrary kick $\delta w_x$ equal $0$,
\be
\begin{split}
\mathbb{P}_{\delta w_x} (S_{r_1} =0&,S_{r_2}=0)\\
&= \exp \left( \int_{x \in \mathbb{R}^d} \delta w_x \,\tilde{u}_{r_1,r_2}( x ) \right)\ .
\label{towpointsdistrib}
\end{split}
\ee
We further restrict  for simplicity to the massless case, i.e.\ without the linear
term $\tilde u_x$  in Eq.~(\ref{InstantonTwoSources}). 
One easily sees from  the latter equation 
that $\tilde{u}_{r_1,r_2}$ takes the scaling form
\be
\tilde{u}_{r_1,r_2}(x) = \frac{1}{(r_1-r_2)^2}\, f\!\left(\frac{2x -r_1 - r_2}{2(r_2-r_1)} \right)\ .
\ee
The function $f(x)$ is   solution of
\be 
f''(x) + f(x)^2 = 0 \ .
\ee  
It diverges at $x=\pm \frac{1}{2}$, vanishes at $x \to \pm \infty$
and is negative everywhere: $f(x) \leq 0$. As  $\delta w_x \geq 0$, the  latter is a necessary condition s.t.\ the probability (\ref{towpointsdistrib}) is bounded by one. 

In the interval $x \in ]-\frac{1}{2},\frac{1}{2}[$, the scaling function $f(x)$ can be expressed 
in terms of the Weierstrass $\mathcal{P}$-function, see (\ref{soluP}),
\be 
f( x ) = - 6 \,\mathcal{P} \!\left( x + \frac{1}{2};g_2=0;g_3
=  \frac{\Gamma(1/3)^{18}}{(2 \pi)^6}  \right)\ .
\label{towSourceSol1}
\ee
The value
of $g_3>0$ is consistent with the required period $2 \Omega=1$, see \eqref{halfper}. 
Note from Appendix \ref{app:W} that there is another solution of the   form 
(\ref{towSourceSol1}) with $g_3 = - \Big( 2 \sqrt{\pi} \frac{\Gamma(1/3)}{4^{\frac{1}{3}} \Gamma(5/6)} \Big)^6<0$
which   violates the condition $f(x) \leq 0$, hence is discarded.
For $|x| \geq 1/2$,  the function $f(x)$ reads 
\be 
f( x ) = - \frac{6}{(|x|-1/2)^2}\ .
\label{towSourceSol2}
\ee
One property of the solution $\tilde{u}_{r_1,r_2}(x)$ is that it diverges as $\sim (x-r_{1,2})^{-2}$ when $x \approx r_{1,2}$.
There are thus two cases:

(i) - the driving $\delta w_x$ is non-zero at one of these points, or vanishes too slowly
near this point (e.g.\ only linearly or slower). Then the integral in \eqref{towpointsdistrib} 
is not convergent, equal to $- \infty$, which implies
$$\mathbb{P}_{\delta w_x} (S_{r_1} =0,S_{r_2}=0) = 0\ .$$
This means that the avalanche contains surely at least one of the points $r_1$ or $r_2$.

(ii) - If $\delta w_x$ vanishes fast enough, for example if $\delta w_x$ is localised away from $x = \pm r_{1,2}$ (e.g $\delta w_x = \delta w \delta (x-y)$ for some $y\in \mathbb{R} \backslash \{r_1,r_2 \} $), the probablity \eqref{towpointsdistrib}  becomes non trivial.

\subsection{Avalanche extension with a local kick}

We now consider  a local kick 
centered at $x=0$, i.e.\ $w_x = \delta w_0 \,\delta(x)$. 
If further $0 < r_1 < r_2$, then
\be
\mathbb{P}_{\delta w_0}\!\left(S_{r_1} =0,S_{r_2}=0 \right) = \mathbb{P}_{\delta w_0}\!\left(S_{r_1} =0\right)\ .
\ee
This comes from the fact that in the interval $x \in [ - \infty,r_1]$, the solution $\tilde{u}_{r_1,r_2}(x)$ 
is   identical to the instanton solution with only one infinite source at $r_1$ (in other word, it does not ``feel" the source in $r_2$). This shows for instance that the support of the avalanche is larger or equal than the set of points
where the driving is non-zero. 

This property also shows that avalanches are connected, i.e.\ it is impossible to draw a plane where the interface did not move  between two moving parts of the interface.
As a function of $r$ (which is one-dimensional),   the support (i.e.\ the set of
points where $S_r >0$) of an avalanche following a local kick at $x=0$ must
be an interval. Since this interval   contains $x=0$ we will write it as $[-\ell_1, \ell_2]$
with $\ell_1>0$ and $\ell_2>0$. This allows to define the extension of an avalanche
as $\ell = \ell_1 + \ell_2$.


To calculate the joint PDF of $\ell_1$ and $\ell_2$ for a kick at $x=0$ 
we consider \eqref{towpointsdistrib} with $r_1=-x_1 < 0 < r_2=x_2$. Using the previous results about the instanton 
equation with two sources, and the fact that the interface model is translationaly invariant, we   obtain the 
joint cumulative distribution for $\ell_1>0$ and $\ell_2>0$:
\be
F_{\delta w_0} (x_1,x_2) : =  \mathbb{P}_{\delta w_0} \left( \ell_1 <x_1, \ell_2 <x_2 \right)\ .
\ee 
It can, for any $x_1,x_2>0$,  be expressed in terms of the function $f$ obtained
in the preceding section, 
\be
\begin{split}\label{cumul_distrib_extension}
F_{\delta w_0} (x_1,x_2) & =
\mathbb{P}_{\delta w_0}\!\left(S_{r_1} =0,S_{r_2}=0 \right)  \\
& = \exp\!\left(\int_x \delta w_0 \delta(x)\,\tilde{u}_{-x_1,x_2}(x)\right)\\
& = e^{\delta w_0 \frac{1}{(x_1+x_2)^2} f \left(- \frac{x_2-x_1}{2(x_1+x_2)}\right)}
\end{split}\ .
\ee
Since the argument of $f$ is within the interval $]- \frac{1}{2},\frac{1}{2}[$ 
we must use the expression \eqref{towSourceSol1}.

From this one can obtain several results. First taking $x_2 \to \infty$
one obtains the PDF of $\ell_1$ alone,
\be
\mathbb{P}_{\delta w} \left( \ell_1 \right) = \frac{12 \delta w}{\ell_1^3} e^{- \delta w \frac{6}{\ell_1^2}} \ .
\ee
A similar result holds  for $\ell_2$. 

\begin{figure}[t]
	\Fig{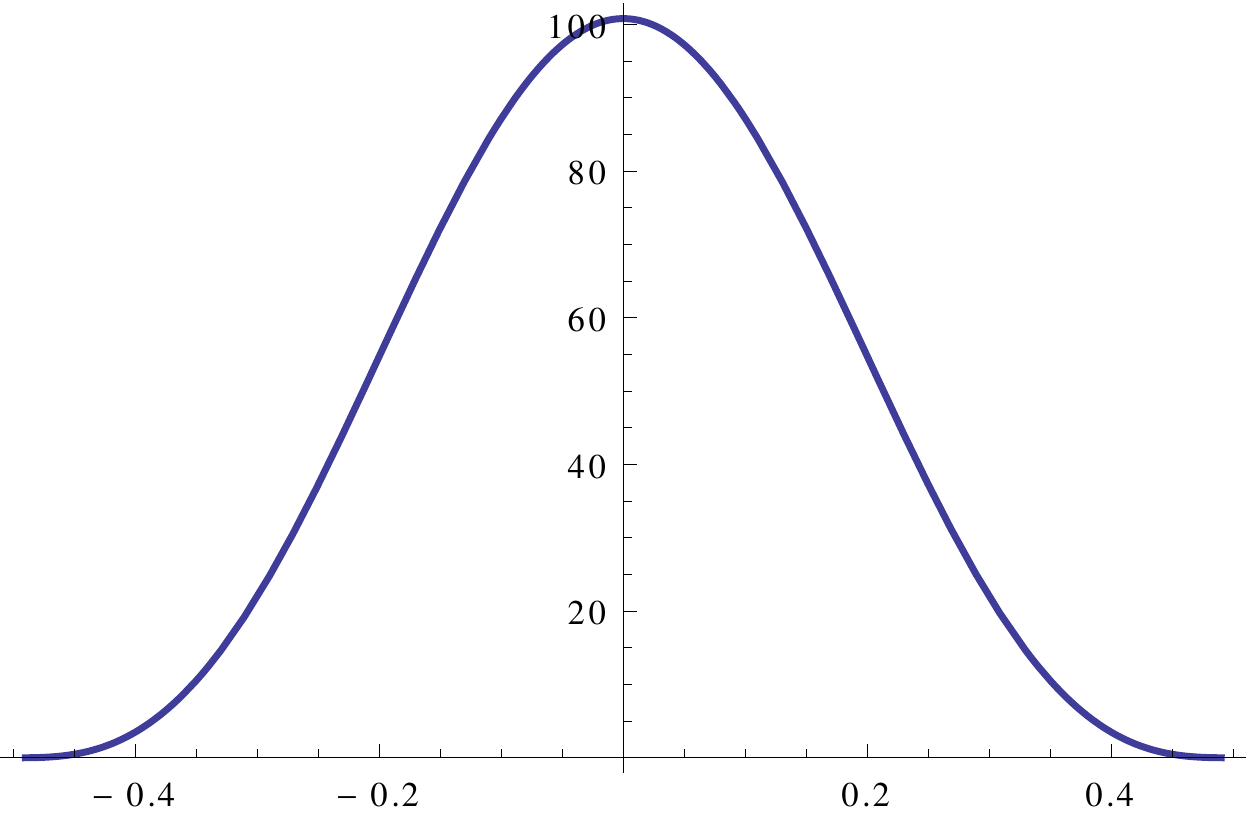}
	\caption{Decay amplitude $R(k)$ as a function of the aspect ratio $k$ involved in the joint density of $\ell$ and $k$, 
	and defined in Eqs.~(\ref{defR1}) and (\ref{defR2}).}
	\label{R(k)}
\end{figure}

In principle, one can now obtain the distribution of avalanches extensions
\be
\mathbb{P}_{\delta w_0}\!\left(\ell\right) = \int_0^{\infty}\!\!\!\! d \ell_1 \int_0^{\infty}\!\!\!\! d\ell_2\,  \delta(\ell - \ell_1 - \ell_2) \partial_{\ell_1} \partial_{\ell_2}
F_{\delta w_0} (\ell_1,\ell_2)
\ee
It has  a rather complicated expression. Let us define in addition to the total length, the aspect ratio
\be
k =  \frac{\ell_1-\ell_2}{2(\ell_1+\ell_2)}  \quad , \quad - \frac{1}{2} < k < \frac{1}{2}\ .
\ee
Using a change of variables, we   obtain the joint density of   total extension and aspect ratio
in the limit $\delta w_0 \to 0$,
\bea\label{defR1}
\rho\left(\ell,k\right) &:=& \lim\limits_{\delta w_0 \to 0} \frac{1}{\delta w_0} \mathbb{P}_{\delta w_0} \left(\ell,k\right)  = \frac{R(k)}{\ell^3} \ ,\; \\
\label{defR2}
R(k) &:=&  6 f(k) + 6 k f'(k) + \left(k^2 - \frac{1}{4}\right) f''(k) \ .~~
\eea
The function $f(x)$ was defined in Eq.~\eqref{towSourceSol1}. While the probability as a function of $\ell$ decays as  $\ell^{-3}$, the  dependence on the aspect ratio is more complicated and plotted in figure \ref{R(k)}. Note that in this expression
$f(k)$ can be replaced by $f_{\rm reg}(k) :=f(k) +  \frac{6}{(k+\frac{1}{2})^2} +  \frac{6}{(k-\frac{1}{2})^2}$,
which    is  a regular function of $k$,  vanishing at $k=\pm \frac{1}{2}$.
Integration over $k$ gives
\bea \label{density1} 
 \rho\left(\ell\right)  &= &\frac{B}{\ell^3} \qquad \text{ with } \\
  B&=&24 + 2 \int_{-1/2}^{1/2} f_{\rm reg}(k)  =8 \sqrt{3} \pi \ .  
\eea

\subsection{Avalanche extension with a uniform kick}

If a kick   extends over the whole system, as e.g.\ a uniform kick $\delta w_x = \delta w$, the avalanche will have almost surely an infinite extension as the local size is non-zero everywhere, 
\be
\mathbb{P}_{\delta w} \left(S_{ r} =0 \right) = 0 \  \text{ for any } \,{ r} \in \mathbb{R}\ .
\ee
However,  in the limit of a small $\delta w$ which is also the limit of a ``single avalanche", we can recover the result for the distribution of extensions. This is consistent with the idea that ``single avalanches" do not depend on the way they are triggered. These calculations allow to obtain the extension distribution without solving explicitly the instanton equation.  (The use of elliptic integrals is in fact equivalent to the use of Weierstrass functions as solutions of the instanton equation, c.f.\ Appendix \ref{app:W}).

We now focus on the following ratio of generating functions
\be
{\langle e^{\lambda_1 s_0 +\lambda_2 s_r}\rangle \over \langle e^{\lambda_1 s_0}\rangle\langle e^{\lambda_2 s_r}\rangle}
\label{generating_ratio}
\ee
in the limit $\lambda_1,\lambda_2 \rightarrow - \infty$. It compares the probability that both local sizes $s_0:= S_0$ and $s_r := S_r$ are simultaneously $0$ to the product of the two probabilties that each one is $0$.

We can express this ratio, using   the instanton-equation approach, as
\be
\begin{split}\label{ratio} 
\lim_{\lambda_1,\lambda_2 \rightarrow - \infty} &{\langle e^{\lambda_1 s_0 +\lambda_2 s_r}\rangle \over \langle e^{\lambda_1 s_0}\rangle\langle e^{\lambda_2 s_r}\rangle}\\
= \exp \!\bigg(&\int_x \delta w_x \Big[\tilde{u}_r(x)-\tilde{u}_{\infty}(x)-\tilde{u}_{\infty}(x-r) \Big]\bigg) 
\end{split}
\ee
where $\tilde{u}_r:=\tilde{u}_{r_1=0,r_2=r}$. We denote by $\tilde{u}_{\infty}:=\tilde{u}_{r_1=0,r_2= \infty}$, the solution of the instanton equation with one source at $r=0$ and the other one at infinity. It is the same as the solution for only one source in $r=0$. The
above expression is valid for any form of driving $\delta w_x$.

\begin{figure}[t]
\Fig{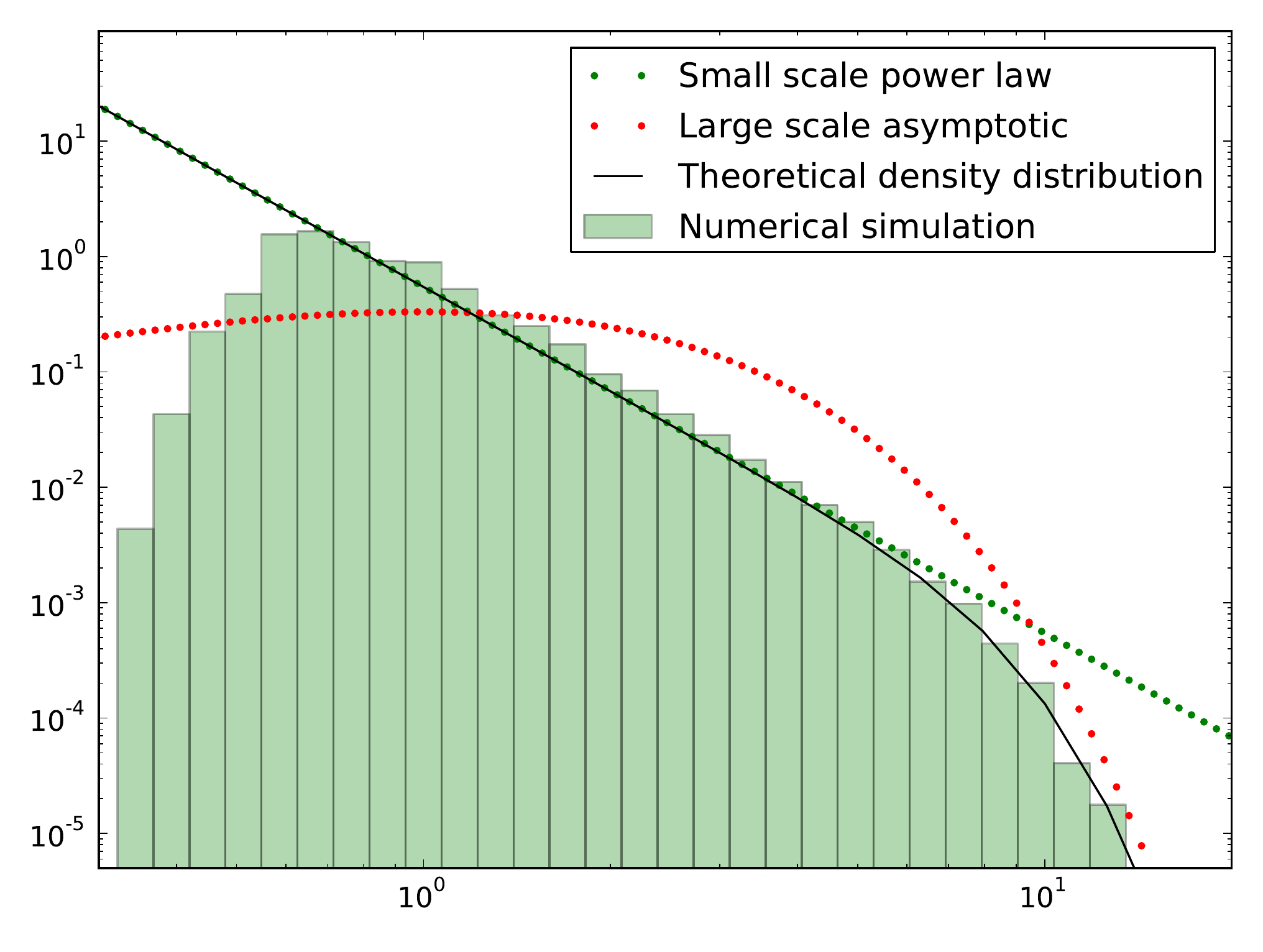}
\caption{The distribution of extensions $\rho(\ell)$, as obtained from the elliptic integrals (\ref{F7}) and (\ref{F8}) (black line). The (straight) green dotted line is the small-$\ell$ asymptotics (\ref{68}), whereas the (curved) red dotted line is the large-$\ell$ asymptotics (\ref{70}). The numerical simulation (green histogram) is cut at small scale due to discretization effects.}
\end{figure}
We can now specify to the case of small and uniform driving $\delta w_x = \delta w$; the quantity of interest is then
\be \label{GeneratingExtension}
Z(r)=\int_x \tilde{u}_r(x)-\tilde{u}_{\infty}(x)-\tilde{u}_{\infty}(x-r) \ .
\ee
While $\tilde{u}_r(x)$ is not integrable, $Z(r)$ is well defined as the two $\tilde{u}_{\infty}$ terms cancel precisely the two non-integrable poles located at $x=0$ and $x=r$.

Using that $\tilde{u}_r$ is a solution of Eq.~\eqref{InstantonTwoSources}, we can obtain an expression of $Z(r)$ as an elliptic integral, see   Appendix \ref{app:elliptic} for   details of the calculation. The formulas  written there are for the massive case, but 
 only allow to get an implicit expression for $Z(r)$.
They   however allow us  to extract the small-scale behavior of the avalanche-extension distribution
(equivalently the massless limit). For small $r$, the behavior of $Z(r)$ is
\be
Z(r) \simeq  \frac{4 \sqrt 3 \pi } {r}  \ .
\ee
To understand the connection with the avalanche extension, we need to get back to the interpretation of (\ref{generating_ratio}). Now that we have specified the kick to be uniform, the two averages of the denominator are independant of $r$, and act only as a normalization constants. The numerator, in the limit of $\lambda_{1,2} \rightarrow - \infty$, is the probability that both $s_0$ and $s_r$ are simultaneously equal to $0$. Deriving this two times \textit{w.r.t.} $r$ (which lets the denominator invariant) gives the probability that the avalanche start in $x=0$ and end in $x=r$. Dividing by $\delta w$ and taking the limit\footnote{Note that the denominators can then be set to unity. There is no ambiguity since the calculation could be performed first at finite but large $\lambda_i$, and setting $\delta w$ to zero after taking the derivative and dividing by $\delta w$, and only at the end taking the limit of infinite $\lambda_i$.} $\delta w \to 0$ , we
obtain the extension density in the limit of a single avalanche as
\bea\label{68}
\rho(\ell) &= &\frac{1}{\delta w} \partial_r^2 e^{\delta w Z(r)}|_{\delta w=0^+,r=\ell} \\
&=& \partial_r^2 \left.\tilde{Z}(r) \right\rvert_{r=\ell} \simeq \tilde B  \ell^{-3}\,\text{ when }\,\ell \rightarrow 0
\nn
\eea
with
\be
\tilde B 
= 8 \sqrt{3} \pi \ .  
\ee
We recover here the  $\ell ^{-3}$ divergence   for small $\ell$ of the extension of  avalanches. 
Note that this calculation gives exactly the same prefactor as in Eq.~(\ref{density1}), which confirms
that we are studying the same object, namely a ``single avalanche". 

Finally,  in the massive case,  one can also compute   the tail of 
the extension distribution, resulting into  (see Appendix \ref{app:elliptic})
\be \label{70}
\rho(\ell)  \simeq 72  \, \ell e^{-\ell} \text{ when } \ell \rightarrow \infty \ .
\ee

\section{Non-stationnary dynamics in the BFM}
\label{sec:non-stat} 
The easiest way to construct a position theory equivalent to the BFM model  define in Eq.~(\ref{BFMdef}) is to consider the non-stationnary evolution of an elastic line in some specific quenched disorder,
\be
\eta \partial_t u_{xt}= \nabla_x^2 u_{xt} + F\left(u_{xt},x\right) + m^2(w_{xt}- u_{xt})\ .
\ee
Here the disorder has the correlations of independent one-sided Brownian motion
\be \label{nonstatcorr} 
\overline{F(u,x)F(u',x')}=2 \sigma \delta^d(x-x')\min(u,u')\ .
\ee
Consider the initial condition $u_{xt=0}=0$. We can then compute the correlation function of the position $$u_{xt}=\int_0^t \dot{u}_{xs}\,ds$$ for a 
uniform   driving $w_t = v t\, \theta (t)$, starting at $t=0$. 
The calculation is sketched in   Appendix \ref{app:nonstat}. In dimensionless
units and in Fourier space, the result reads
\bea \label{nonstatBFM}
\langle u_{qt}u_{-qt} \rangle^c &=& v \Bigg[ \frac{2 q^2 (t-1)+2 t -5}{\left(q^2+1\right)^3}-\frac{4 e^{-\left(q^2+1\right) t}}{q^2 \left(q^2+1\right)^3} \nn\\
&&+\frac{4 e^{- t}}{q^2 \left(2 q^2+1\right)}+\frac{e^{-2 \left(q^2+1\right) t}}{\left(q^2+1\right)^3 \left(2 q^2+1\right)} \Bigg]\ .~~~~~~~~
\eea
At large times, the displacement correlations behave as (restoring units)
\be \label{correl} 
\langle u_{qt} u_{-qt} \rangle^c \underset{ t \rightarrow \infty}{\simeq} \frac{2 \sigma v  t }{(q^2+m^2)^2}\ .
\ee
The $q$ dependence is similar to the so-called Larkin random-force model \cite{Larkin1970},
but with a time-dependent amplitude, i.e.\ the effective disorder is
growing with time, which is natural given the correlations \eqref{nonstatcorr}. The correlation of the position
thus remains non-stationary at all times\footnote{Note that there are
stationary versions of the BFM, which we will not discuss here, see
discussions in e.g.\ \cite{LeDoussalWiese2011a,DobrinevskiLeDoussalWiese2011b,LeDoussalWiese2012a}.}.

From Eq.~(\ref{correl}) one obtains the correlations of the displacement in real space, still in the large-$t$ limit
\bea
\overline{ (u_{xt}-u_{0t})^2 } && \simeq 2 v t \int \frac{d^dq}{(2 \pi)^d} \frac{1}{(q^2+m^2)^2} (1-\cos q x) \nn \\
&& \sim v t \times x^{2 \zeta_L} 
\eea 
with $\zeta_L= (4-d)/2$ the Larkin roughness exponent. Note that the
average displacement is
$\overline{u_{xt}} = v t - \frac{1-e^{-m^2 t}}{m^2}$ (see Appendix \ref{app:nonstat} ).
Hence we see that the BFM roughness scaling   $u \sim x^{4-d}$
is dimensionally consistent with the
correlation at large  times, 
\be
\overline{ (u_{xt}-u_{0t})^2 }  \simeq 2 ~ \overline{u_{xt}} ~ x^{4-d} \ .
\ee
This result, $\zeta = 4 -d = \varepsilon$, is in agreement with the FRG approch: 
the position theory of the BFM model is an exact fixed point for the flow 
equation of the FRG with a roughness exponent $\zeta = \varepsilon$, 
as discussed in \cite{LeDoussalWiese2011b,DobrinevskiLeDoussalWiese2011b}.

\section{Conclusion}

We presented a general investigation of the Brownian Force Model, using its exact solvability   via the instanton equation in various settings. After reviewing the results and the calculations of \cite{LeDoussalWiese2008c,LeDoussalWiese2011a,DobrinevskiLeDoussalWiese2011b,LeDoussalWiese2012a}, 
we extended the study in several directions.
	
First, we  computed   observables containing    information about the spatial structure of avalanches in the BFM: the joint density of $S$ and $S_0$ (or  equivalently, the distribution of the local size 
$S_0$ at   fixed total global size $S$), and the distribution of the extension $\ell$ of an avalanche. 
These distributions display power laws in their small-scale regime, 
which we     recovered using scaling arguments, together with universal amplitudes. 

We also extended the method to study new driving protocols relevant to distinct experimental setups.
The derived results show new exponents for the small-scale behavior of the global avalanche-size distribution following a locally imposed displacement, and for the small-scale behavior 
of the local-size distribution following a localized kick.

Finally, we presented results for the non-stationary dynamics of the BFM, focusing
on observables which exist only in the position theory, such as the roughness exponent. 
This   explains why both the Larkin   roughness and the BFM roughness (emerging from the FRG approach),
play a role in this model,     depending on whether  the driving  is stationary or not.

\acknowledgements
We thank A.\ Rosso, A.\ Kolton and A.\ Dobrinevski for stimulating discussions,   PSL for support by Grant No.\ ANR-10-IDEX-0001-02-PSL, as well as KITP for hospitality and support in part by NSF Grant No.\ NSF PHY11-25915.

\appendix

\section{Airy functions} 
\label{app:airy} 
We recall the definition of the Airy function:
\be  \label{airydef}
\text{Ai}(z) := \int_{-\infty}^{\infty} \frac{dt}{2 \pi} e^{i \frac{t^3}{3} + i z t}\ .
\ee 
The following  formula is usefulfor $a \in \mathbb{R}^*$,
\bea \label{Phi}
 \Phi(a,b,c) &=& \int_C \frac{dz}{2 i \pi} e^{a \frac{z^3}{3} + b z^2 + c z} \\
& =& |a|^{-1/3} e^{\frac{2 b^3}{3 a^2} - \frac{b c}{a}} 
\text{Ai}\!\left(\frac{b^2}{|a|^{4/3} } - \frac{c ~ {\rm sgn}(a)}{|a|^{1/3}} \right)\ . \nn
\eea 
It can be obtained from (\ref{airydef}), deforming the  contour $C$, e.g.\ to $z=- \frac{b}{a} + i \mathbb{R}$.

\section{General considerations on the instanton equation}
\label{AppendixB}

\subsection{Sourceless equation} 

\subsubsection{Massive case} 

\begin{figure}[t]
\Fig{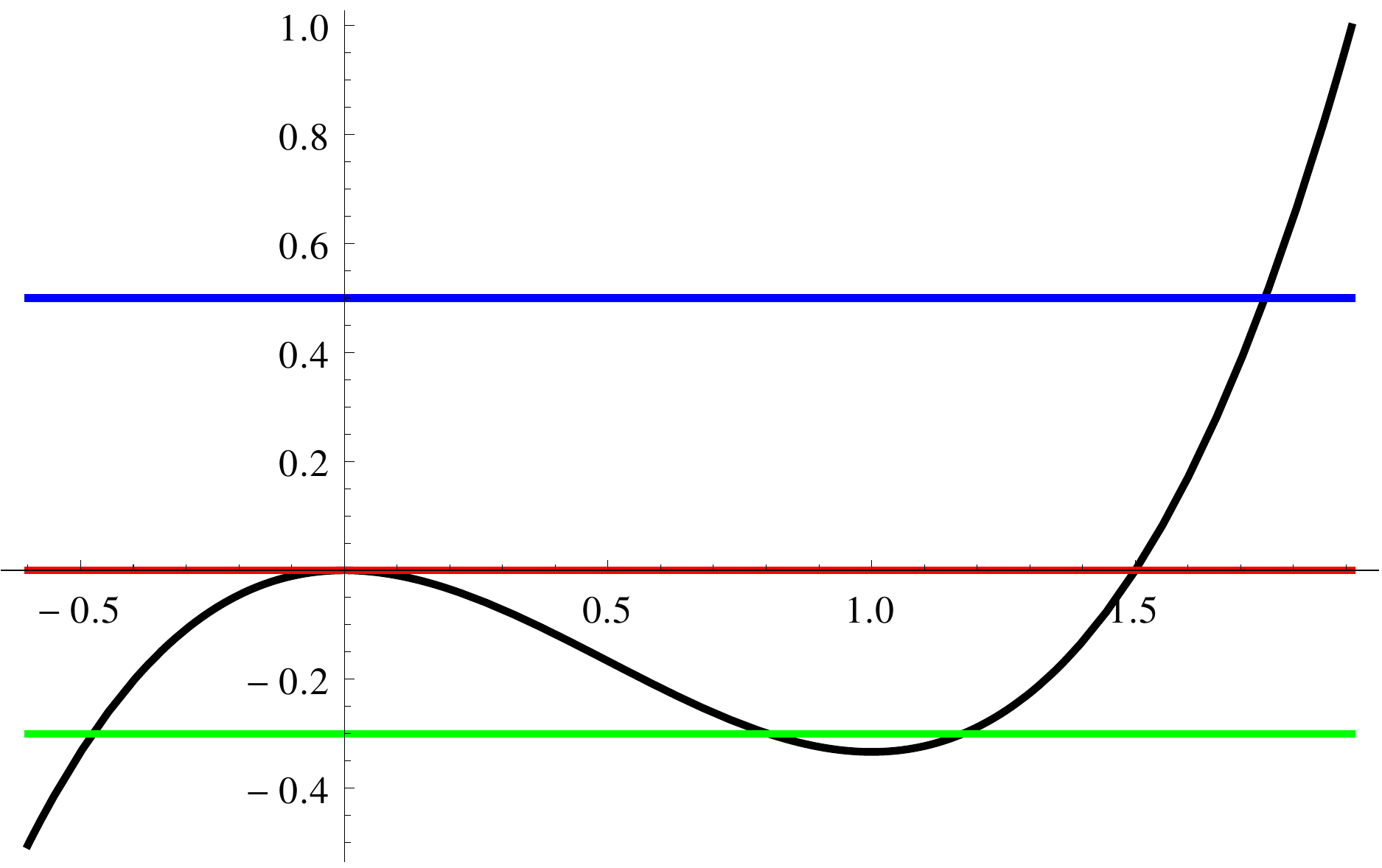}
\caption{Representation of the potential energy $V(y)$ as a function of $y$, and lines of constant total energy, with $E=0$ in red, $E>0$ in blue and, $E<0$ in green.}
\label{energyPlot}
\end{figure}

\begin{figure}[t]
\Fig{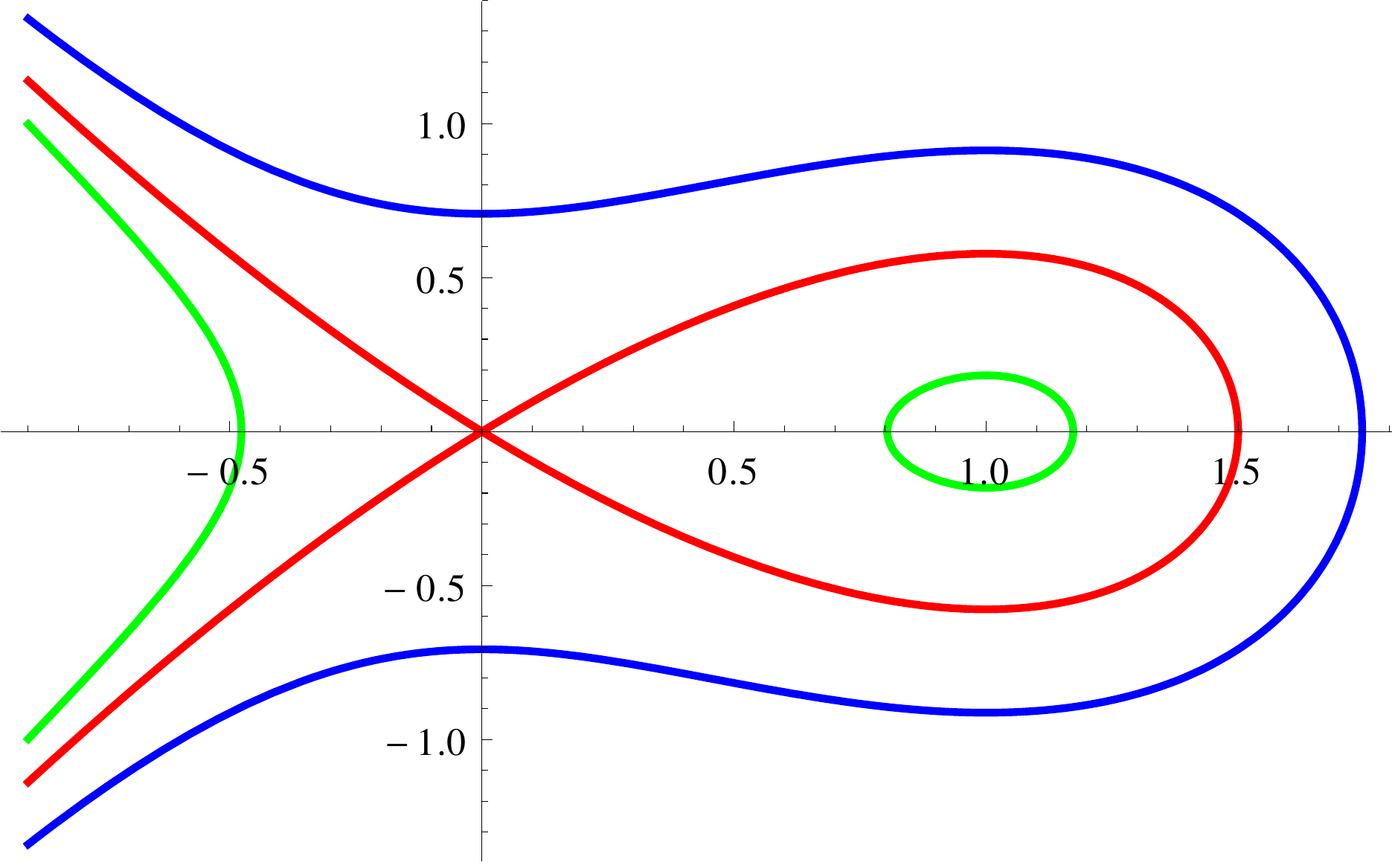}
\caption{Phase-space diagram, i.e.\ trajectories represented with $y'$ as a function of $y$. The case $E=0$ is in red, $E>0$ in blue and and $E<0$ in green. We can see that properties of the solution (periodicity, divergences, etc.) strongly depend on the value of $E$.}
\label{PhaseDiag}
\end{figure}

It is useful to start with the simpler sourceless instanton equation
\begin{equation}\label{E}
y'' = y -y^2\ .
\end{equation}
Here we denote by a prime the derivative with respect to $x$.
It can be interpreted as the classical equation of motion of a particule (of mass $2$) in a potential {$V(y) = - y^2+{2 y^3 \over 3}$}, represented in Fig.\ \ref{energyPlot}. Multiplying by $y'$ and integrating once, we obtain 
$y' = \pm \sqrt{E - V(y)}$, where $E$ is a real integration constant equivalent to the total  ``energy'' of the particle.
Its phase-space diagram $(y,y')$   is represented in Fig.~\ref{PhaseDiag}.

From figures \ref{energyPlot} and \ref{PhaseDiag}, we see that:
\begin{figure*}[t]
\Fig
{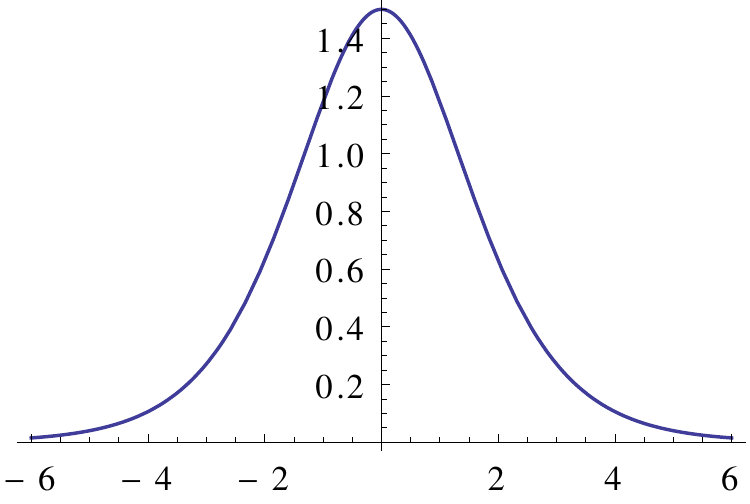}
\Fig
{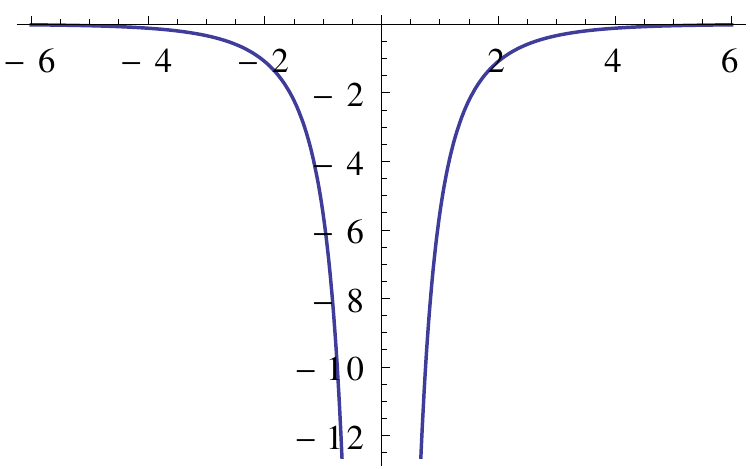}
\caption{Solutions with  energy $0$ of   equation \eqref{E}; left : $y^+(x)$, right : $y^-(x)$.}
\label{solutionPlot}
\end{figure*}

(i) - there is exactly one positive $E=0$ solution $y^+(x)$ 
defined for all $x\in \mathbb{R}$, up to a shift $x \to x+x_0$. It reads
\be \label{y+}
\begin{split}
&\int_{y^+(x)}^{3/2} \frac{dy}{\sqrt{y^2 - \frac{2}{3} y^3}} = |x| \\
&\Leftrightarrow\; y^+\!(x) = \frac{3}{1 + \cosh x} = \frac{3}{2} \bigg[1 - \tanh^2\!\left(\frac{x}{2}\right)\bigg]\ .
\end{split}
\ee

(ii) - There is exactly one negative $E=0$ (zero energy) solution $y^-(x)$ defined for all $x\in \mathbb{R}^*$,
namely
\be\label{y-}
\begin{split}
&\int^{y^-\!(x)}_{-\infty}\!\!\!\frac{dy}{\sqrt{y^2 - \frac{2}{3} y^3}} = |x| \\
&\Leftrightarrow\; y^-(x) = \frac{3}{1 - \cosh x} = \frac{3}{2} \bigg[1 - \coth^2\!\left(\frac{x}{2}\right)\bigg]\ .
\end{split}
\ee

(iii) - There are two classes of solutions with $E \neq 0$. The first class is
defined   on an interval of finite length $r(E)$ with
\be
r(E)=2 \int^{t}_{-\infty} \frac{dy}{\sqrt{E + y^2 - \frac{2}{3} y^3}} 
\ee
where $t\neq 0$ denotes the smallest real root of $E=-t^2 + \frac{2}{3} t^3$.
This integral is convergent at large negative $y$ due to the cubic term, 
and also convergent near the root $y=t$ (for $E \to 0$ it diverges logarithmically). 
If one chooses $x=0$ as   center of the interval, the
solution $y(x)$ satisfies
\be
\int^t_{y(x)} \frac{dy}{\sqrt{E + y^2 - \frac{2}{3} y^3}} = |x|\ .
\ee
It diverges at both ends $x = \pm r(E)/2$. It is sometimes more 
convenient to choose $x=0$ as the endpoint of the interval $]0,r(E)[$.
Then, for $x \in ]0,r(E)[$ one has
\be
\int^{y(x)}_{-\infty} \frac{dy}{\sqrt{E + y^2 - \frac{2}{3} y^3}} = x\ .
\ee
Setting $y = \frac{1}{2} - z$, this can be rewritten as
\be
\sqrt{6} \int_{\frac{1}{2} - y(x)}^{\infty} \frac{dz}{\sqrt{4 z^3 - 3 z + (1+6 E)}} = x\ .
\ee
This gives, in terms of the Weirstrass elliptic function $\cal P$,
\be
y(x) = \frac{1}{2}- {\cal P}\!\left(\frac{x}{\sqrt{6}} ;g_2=3,g_3=-1- 6 E\right)\ .
\ee
It diverges at $x=0$ and $x=r(E)$, and is the proper solution
on the   interval $]0,r(E)[$, see Appendix \ref{app:W}. 

The second class of solutions with $E \neq 0$ exists only for $- \frac{1}{3} < E <0$; these solutions 
are periodic on the whole real line. As can be seen from Figs.\ \ref{energyPlot} and \ref{PhaseDiag}, 
$y(x)$ varies in a bounded and strictly positive interval. We will not discuss these solutions
as they will not be   needed below.

\subsubsection{Massless case} 

Consider now the massless sourceless equation,
\be
y'' = - y^2\ .
\ee
The analysis is similar to the massive case discussed above with $V(y)= - \frac{2}{3} y^3$. Its solutions have the following properties: 

(i) - there is no  positive $E=0$ solution.

(ii) - There is only one negative $E=0$ solution $y^-(x)$ defined for all $x\in \mathbb{R}^*$, 
\be
\int^{y^-(x)}_{-\infty} \frac{dy}{\sqrt{- \frac{2}{3} y^3}} = |x|  \Leftrightarrow y^-(x) = - \frac{6}{x^2} \ .
\ee 
It can be  obtained by considering the limit $x \ll 1$ in the solution (\ref{y-}). 

(iii) - There is now only one class of solutions with $E \neq 0$ (the periodic ones
have disappeared). They are defined   on an interval of   length $r(E)$.
They have $E= \frac{2}{3} t^3$,  hence $t=(3 E/2)^{1/3}$ and
\bea
r(E)&=& 2 \int^{t}_{-\infty} \frac{dy}{\sqrt{\frac{2}{3} t^3 - \frac{2}{3} y^3}}\nn \\
&=& \begin{cases}
\sqrt{6 \pi} \left(\frac{2}{3 |E|}\right)^{\!1/6} \frac{\Gamma(1/3)}{\Gamma(5/6)}\, , \quad E>0  \\
\sqrt{6 \pi} \left(\frac{2}{3 |E|}\right)^{\!1/6} \frac{2 \Gamma(7/6)}{\Gamma(2/3)}\, , \;\; E<0\ .
\end{cases}
\eea
The solution $y(x)$ 
satisfies for $x \in ]0,r(E)[$
\be
\int^{y(x)}_{-\infty} \frac{dy}{\sqrt{E - \frac{2}{3} y^3}} = x\ .
\ee
It can be expressed in terms of the Weirstrass function, 
\be
y(x) = - {\cal P}\!\left(\frac{x}{\sqrt{6}} ;g_2=0,g_3= - 6 E\right)\ .
\ee
It diverges at $x=0$ and $x=r(E)$. The periods are
consistent with $\sqrt{6} \times 2 \Omega$ (see Appendix \ref{app:W}) 
using the relation $\frac{\Gamma(7/6)}{\Gamma(2/3)} = \frac{\Gamma(1/3)^3}{4\times 2^{1/3} \pi^{3/2}}$. 
Note also the relation $\frac{\Gamma(1/3)}{\Gamma(5/6)} =\frac{2\times 2^{2/3} \pi^{3/2}}{3 \Gamma(2/3)^3}$.

\subsection{Instanton solution with a single delta source}

We now use these results to construct the solutions 
in presence of sources. For a single delta source this was done in \cite{LeDoussalWiese2008c}
and \cite{LeDoussalWiese2012a}. We first recall and then extend this analysis, as a   more
general approach is needed here.

\subsubsection{Massive case} 
\begin{figure}[t]
\Fig{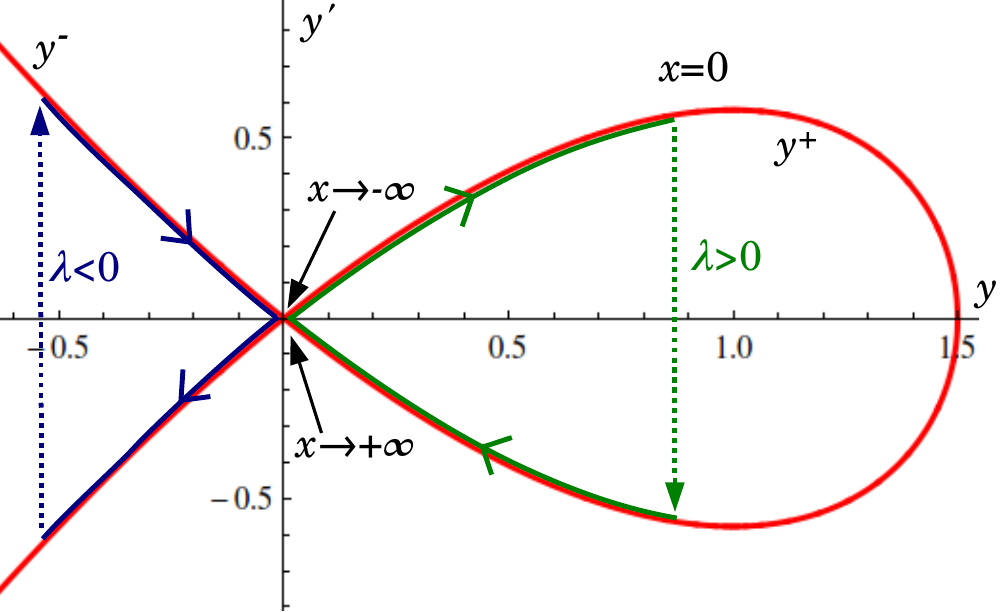}
\caption{Graphical representation of the construction of solutions of the instanton equation for $\lambda > 0$ (blue) and $\lambda < 0$ (green). The dotted part of the curve represents the discontinuity in the derivative. The red line represents the $E=0$ solution of \eqref{E}, the only one needed to solve the instanton equation with one local source.}
\label{GraphicalInstanton}
\end{figure}

Consider   the instanton equation
\be  \label{inst1} 
\tilde u''(x) - \tilde u(x) + \tilde u(x)^2 = - \lambda \delta(x)\ . 
\ee  
We are looking for a solution defined for all $x \in \mathbb{R}$. Other physical requirements\footnote{Because of finite range elasticity, the
the effect at $x=0$ of a kick at $x$ must decay at large $x$. Because of the cutoff $S_m$, the
positive integer moments of avalanche sizes must exist} (e.g.\ from the derivation of the
dynamical action) is that $\tilde u(x)$ vanishes as $x \to \pm \infty$, and that 
the solution is analytic around $\lambda=0$ (obtainable in a power series in $\lambda$).
We need a function  which is piecewise solution of Eq.~\eqref{E} for $x \in ]-\infty,0[$ and for $x \in ]0,\infty[$,  with a
discontinuity in its derivative,
\be  \label{disc} 
\tilde u'(0^+) - \tilde u'(0^-)  = - \lambda \ .
\ee  
As we have seen in  the previous section, in order to be  defined on an infinite interval, 
it must be constructed from the zero-energy $E=0$ solutions $y^\pm(x)$ of \eqref{E} up to a shift $x \to x+ x_0$. 
By symmetry it reads $\tilde{u}(x)=y^{\pm}(|x| + x_0) $ where $x_0\equiv x_0(\lambda)$ is chosen to satisfy the condition (\ref{disc}).
The procedure is illustrated in Fig.\ \ref{GraphicalInstanton}. Note that the sign of $\lambda$ dictates which of the branches $\pm$ must be chosen.
To summarize, 
\be
\tilde u^\lambda(x) = \frac{3}{1+ s_\lambda \cosh(|x|+x_0)} = \frac{3}{2} \big[1 - h_\lambda(|x|+x_0)^2\big]\ .
\ee
The function $x_0(\lambda)$ is determined from
\be \label{eqla} 
\lambda = \frac{6 s_\lambda \sinh (x_0)}{\big[1+ s_\lambda \cosh (x_0)\big]^2} = \frac{3}{2} h_\lambda(x_0) \big[1 - h_\lambda(x_0)^2\big]
\ee
with $s_\lambda = {\rm sgn}(\lambda)$, $h_\lambda(x)=\tanh(\frac{x}{2})$ for $\lambda>0$ and $h_\lambda(x)=\coth(\frac{x}{2})$ for $\lambda<0$. 
\footnote{Note that formally $x_0 \to x_0 + i \pi$ is equivalent to $\lambda \to - \lambda$.}

\begin{figure}[t]
\Fig{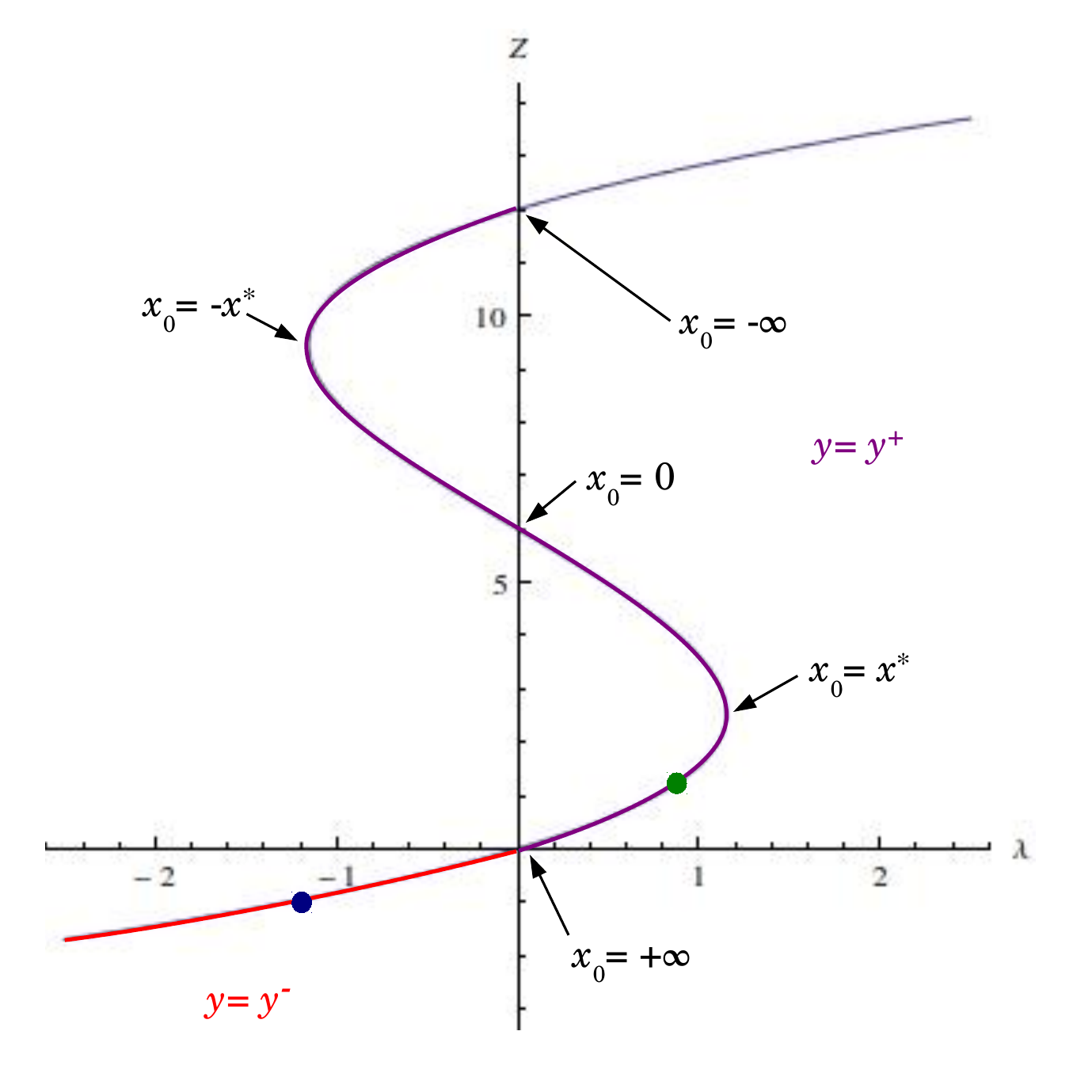}
\caption{The generating function $Z(\lambda)= 6 (1-z)$ is represented here with some indications of the link with the construction of the instanton solution; the green and blue dot correspond to the solutions represented in figure \ref{GraphicalInstanton}.}
\label{Z(lambda)}
\end{figure}

This form   does not make explicit that $\tilde u^{\lambda}(x)$ is analytic in $\lambda$ near $\lambda=0$. We will thus use the
following equivalent form. Introduce $z = h_\lambda(x_0)$. Equation (\ref{eqla}) can then be rewritten as a cubic equation for $z\equiv z(\lambda)$,
\be \label{zeq} 
\lambda = 3 z (1-z^2) \ .
\ee 
The trigonometric addition rules allow to rewrite
\be \label{soluutilde}
\begin{split}
\tilde u^\lambda(x) &= \frac{3 (1-z^2)}{2 \Big[\cosh\!\left(\frac{x}{2}\right) + z \sinh\!\left(\frac{|x|}{2}\right)\Big]^2 } \\
& = \frac{6 (1-z^2) e^{- |x|} }{\big[1 + z + (1-z) e^{-|x|}\big]^2}\ .
\end{split}
\ee
The appropriate branch for (\ref{zeq}) is the one for which $z \to 1$ as $\lambda \to 0$ (corresponding to
$x_0 \to \infty$). As can be seen in Fig.\ \ref{Z(lambda)}, this branch is defined for $\lambda \in ]-\infty, \lambda_c=\frac{2}{\sqrt{3}}[$,  
while $z(\lambda)$ decreases from $z(-\infty)=\infty$ to $z_c=z(\lambda_c)=1/\sqrt{3}$. 
The other branches are solutions of (\ref{inst1}) but   do not
satisfy the physical requirements mentioned above. 

Equations (\ref{zeq}) and (\ref{soluutilde}) thus define the solution to the
instanton equation for $\lambda \in ]-\infty, \lambda_c[$, in a way which is explicitly
analytic around $\lambda=0$. For instance one can check that the small-$\lambda$ expansion
\be
\tilde u^\lambda(x) = \frac{\lambda}{2} e^{-|x|} + \frac{\lambda^2}{6} \left(e^{-|x|} - \frac{1}{2} e^{-2 |x|}\right) + \mathcal{O}(\lambda^3)
\ee 
obtained by iteratively solving  Eq.~(\ref{inst1}) at small $\lambda$, is   reproduced
by Eqs.~(\ref{zeq}) and (\ref{soluutilde}). 

Finally the partition sum corresponding to an homogeneous kick is
 expressed as
\be
Z(\lambda) = \int_{-\infty}^{\infty}\!\!\!dx ~\tilde u^{\lambda}(x) = 6 (1-z)\ .
\ee
Hence, from Eq.~(\ref{zeq}), it satisfies
\be \label{Z_lambda_equation}
\lambda = \frac{1}{72} Z (Z-6) (Z-12)\ ,
\ee
recovering the result obtained in \cite{LeDoussalWiese2008c}. 

\subsubsection{Massless case}

The massless instanton equation
\be \label{inst2} 
\tilde u''(x) + \tilde u(x)^2 = - \lambda \delta(x) 
\ee
is solved similarly. For $\lambda<0$ there is a solution defined for all $x \in \mathbb{R}$, 
\be \label{solumassless} 
\tilde u^\lambda(x) = - \frac{6}{(|x| + x_0)^2}  \quad , \quad x_0^3 = -\frac{ 24}{\lambda}\ .
\ee
Note that for the massless case the physical solution is not required to be analytic in $\lambda$
at $\lambda=0$ (i.e.\ integer moments of avalanche sizes diverge). 
This solution can be obtained from (\ref{soluutilde}) in the (formal) double
limit of small $x$ and large $z$, with $x_0 = 2/z$. The equation determining $z$
  now is $\lambda = -3 z^3$. The generating function for a uniform kick becomes
$Z=- 6 z= (72 \lambda)^{1/3}$. 

\section{Calculation of probabilities and densities of $S_0$}

For an arbitrary kick $\delta w_x$, in the massive case, the Laplace transform of the distribution 
of local size is 
\be \label{lt1} 
\int\!\!dS_0\, e^{\lambda S_0} P_{ \delta w_x  }(S_0) = \exp\!\left(\!L^{d-1} \int\!\!dx\,\delta w_x \tilde u^\lambda(x)\right)\ .
\ee
Here $\tilde u^\lambda(x)$ is given in Eq.~(\ref{soluutilde}). Performing the Laplace inversion 
in general is difficult, but there are some tractable cases. 

\subsection{Uniform kick}

Let us start with a uniform kick $\delta w_x=\delta w$, and $\delta \hat w=L^d \delta w$.
It is more efficient to take a a derivative of Eq.~(\ref{lt1}) w.r.t. $\lambda$ and
write the Laplace inversion   for $S_0 P_{w}(S_0)$,
\be 
S_0 P_{\delta w}(S_0) = \int_C \frac{d \lambda}{2 i \pi} e^{- \lambda S_0} \partial_\lambda e^{6 \delta \hat w  (1-z(\lambda))} \ .
\ee  
Here $C$ is an appropriate contour parallel to the imaginary axis 
and we used that $\int dx\, \tilde u(x) = 6 (1-z)$. The function $z(\lambda)$ is solution of $\lambda = 3 z(1-z^2)$. 
One can now use $z$ as integration variable and rewrite
\be 
S_0 P_{\delta w}(S_0) =  6 \delta \hat w e^{6 \delta \hat w} \int_C \frac{d z}{2 i \pi} e^{-  3 z(1-z^2) S_0} e^{- 6 \delta \hat w z} \ ,
\ee  
using $d\lambda \partial_\lambda = dz \partial_z$.
We will be sloppy here about the integration contour, as this procedure is heuristic to guess the 
result, which will then be tested (see below). As the exponential contains a cubic term,  we 
 use the Airy integral formula of Appendix \ref{app:airy} leading to
\be 
S_0 P_{w}(S_0) =  6 \delta \hat w e^{6 \delta \hat w} \Phi(a,b,c) \ .
\ee  
Here $\Phi$ is defined in Eq.\ (\ref{Phi}), 
with $a=9 S_0$, $b=0$ and $c=-( 3 S_0 + 6 \delta \hat w)$. This immediately leads to
  formula (\ref{LocalDistrib}) in the main text. We have checked numerically that it
reproduces the correct Laplace transform (\ref{lt1}) for $\lambda < \lambda_c$.

\subsection{Local kick}
\label{app:local} 
For a local kick it is possible to calculate the PDF of the local jump at the
position of the kick. 

Consider a local kick at $x=0$, i.e.\ $\delta w_x=\delta w_0 \delta(x)$. 
For simplicity in this subsection we set $d=1$. Inserting this value in (\ref{lt1}) we find that the
LT of the PDF of the local size at the same point $S_0$ reads
\be \label{lt2} 
\int dS_0\,e^{\lambda S_0} P_{\delta w_0}(S_0) = e^{  \frac{3}{2} (1-z^2) \delta w_0} 
\ee
using $\tilde u^\lambda(0)= \frac{3}{2} (1-z^2)$. The same manipulations as above lead to
\bea
S_0 P(S_0)&=& - \int_C \frac{d z}{2 i \pi} e^{-  3 z(1-z^2) S_0} \partial_z e^{\frac{3}{2} (1-z^2) \delta w_0} \\
&=& 3 \delta w_0 e^{\frac{3\delta w_0}{2}  }  \int_C \frac{d z}{2 i \pi} z ~ e^{-  3 z(1-z^2) S_0 - \frac{3}{2} z^2 \delta w_0}\nn \\
&=& 3 \delta w_0 e^{\frac{3\delta w_0}{2}  } \partial_c \Phi(a,b,c)|_{a=9 S_0,b=-\frac{3\delta w_0}{2} ,c=-3 S_0}\nn\ .
\eea
Using Eq.\ (\ref{Phi}) leads to
\bea \label{resloc} 
P_{\delta w_0}(S_0)&=&\frac{ \delta w_0 e^{\delta w_0- \frac{\delta w_0^3}{36 S_0^2}}}{3^{1/3} S_0^{5/3}}   \left[ \frac{\delta w_0}{2\times 3^{1/3}  S_0^{2/3}} \text{Ai}(u) - \text{Ai}'(u)\right] \nn \\
u &=& 3^{1/3} S_0^{2/3} + \frac{\delta w_0^2}{4 \times 3^{2/3} S_0^{4/3}}\ .
\eea 
We can check normalization, and that $ \langle S_0 \rangle= \frac{1}{2} \delta w_0$, consistent with 
  the small-$\lambda$ expansion of (\ref{lt2}). The asymptotics are
\be
P_{\delta w_0}(S_0) \simeq \left\{
\begin{array}{ll}
\dfrac{\delta w_0^{3/2}e^{\frac{\delta w_0}{2}-\frac{\delta w_0^3}{18 S_0^2}}}{\sqrt{6 \pi } S_0^2}\;\; &\text{for }\;S_0 \ll1  \\
\vspace{-0.25cm}\\
\dfrac{ \delta w_0 e^{\delta w_0-\frac{2}{\sqrt{3}} S_0}}{2 \sqrt[4]{3} \sqrt{\pi } S_0^{3/2}}&\text{for }\;S_0 \gg1\ .
\end{array}\right.
\ee
This result, and the new exponent $\tau=5/3$ of the divergence at small $S_0$, which appear when $\delta w_0 \to 0$, is discussed in the main text.

\section{Calculation of the joint density of $S$ and $S_0$}
\label{sec:joint} 

We will obtain the joint density from the generating function of $S_0$ and $S$,  
\be
\langle e^{\lambda S_0 + \mu S} \rangle = e^{ \int_x \delta w_x \tilde{u}_x}
\ee
in terms of the solution of the instanton equation. Let us consider a
uniform kick $\delta w_x=\delta w$. 

\subsection{Instanton equation and its solution} 

\subsubsection{Massive case}
Here $\tilde{u}$ (that we will also denote $\tilde{u}^{\lambda,\mu}$ to make the dependence on the sources explicit) is the solution, in the  variable $x$, of the instanton equation
\be
\tilde{u}''-\tilde{u}+ \tilde{u}^2 =-  \lambda \delta(x) - \mu\ . 
\label{instantonJoint}
\ee
We must solve this equation with similar requirements as
discussed below for Eq.~\eqref{inst1}, except that now the instanton 
goes to a constant at infinity (since the source acts everywhere). 
Clearly, the new uniform source can be removed
by a   shift $\tilde u \rightarrow \tilde u + c$, where the constant $c$ verifies $\mu=c-c^2$. 
This results in the mass term $-\tilde{u} \to -(1-2 c)\tilde{u}$, which can
be brought back to Eq.\ (\ref{instantonJoint}) with $\mu=0$, i.e.\ Eq.\ (\ref{inst1}),
by a simple scale transformation. At the end one can check that given 
$\tilde u^\lambda(x)$ the solution of Eq.\ (\ref{inst1}),   the 
  solution of Eq.\ (\ref{instantonJoint}), noted $\tilde u^{\lambda,\mu}(x)$, is given  by
\be \label{solu1} 
\tilde u^{\lambda,\mu}(x) = \frac{1- \beta^2}{2} + \beta^2 \tilde u^{\lambda/\beta^3}(\beta x)\ .
\ee
The constant  $\beta >0$ such that
\be \label{defBeta}
\beta^2 = \beta_\mu^2 := \sqrt{1- 4 \mu}\ .
\ee
In summary,  the instanton solution is   
\be \label{solu2} 
\tilde u^{\lambda,\mu}(x)  = \frac{1- \beta^2}{2} +  \frac{6 \beta^2 (1-z^2) e^{- \beta |x|} }{\big[1 + z + (1-z) e^{- \beta |x|}\big]^2} \ ,
\ee
where $z$ is the solution of 
\be  \label{eqz}
   \frac{\lambda}{\beta^3} = 3 z (1-z^2)\ .
\ee  
It is connected to $z=1$ at $\lambda=0$. 

\subsubsection{Massless case}
It is useful to also give  the solution in the massless case, for which one needs to solve
\be 
\tilde{u}'' + \tilde{u}^2 =-  \lambda \delta(x) - \mu 
\label{instantonJointMassless}
\ee  
for $\mu \leq 0$. Using a shift and a rescaling we can check that the
solution now is 
\be \label{solu1massless} 
\tilde u^{\lambda,\mu}(x) = \frac{- \beta^2}{2} + \beta^2 \tilde u^{\lambda/\beta^3}(\beta x)\ .
\ee
The parameter $\beta >0$ such that $\beta^2 = \sqrt{- 4 \mu}$, and $\tilde u^{\lambda}(x)$ is the {\it massive} instanton solution. 
In summary, this gives
\be   \label{solu2massless} 
   \tilde u^{\lambda,\mu}(x)  = \frac{- \beta^2}{2} +  \frac{6 \beta^2 (1-z^2) e^{- \beta |x|} }{\big[1 + z + (1-z) e^{- \beta |x|}\big]^2} 
\ee  
where $z$ is again the solution (\ref{eqz}). If $\mu \to 0$, hence $\beta \to 0$ we recover the
massless instanton \eqref{solumassless}. 

%

\medskip

\subsection{Joint distribution} 
Let us again consider  the massive case. 
To obtain the joint probability distribution $P_{\delta w}( S,S_0)$, we need to calculate the generating function $Z(\lambda,\mu)$,
\be
\begin{split}
\langle e^{\lambda S_0 + \mu S} \rangle=&\int_0^{\infty} \int_0^{\infty} P_{\delta w}(S_0,S) 
e^{\lambda S_0 + \mu S} dS \, dS_0\\ =& e^{ \delta w Z\left(\lambda,\mu \right)}
\label{JointProbaDef}
\end{split}
\ee
Integrating \eqref{solu2}, we obtain
\be
\begin{split}
Z(\lambda,\mu)= \int_x \tilde u^{\lambda,\mu}(x) &= L^d \frac{1- \beta^2}{2}
+ L^{d-1} 6 \beta z (1-z) \\
&=: L^d Z_1(\mu)  + L^{d-1} Z_2 \left( \lambda , \mu \right) \ .
\end{split}\label{ZV2}
\ee 
$Z_1(\mu)$ is the generating function for the distribution of the total size of avalanches and $Z_2(\lambda,\mu)$ a new term defined by \eqref{ZV2}. The volume factors come from the coordinates along which the instanton solution is constant.

From equations \eqref{eqz} and \eqref{ZV2}, we can express $\lambda$ as a function of $Z_2$ and $\beta$, 
\be
\lambda= 3 \beta^3 \left( 1- {Z_2 \over 6 \beta} \right) \left[1- \left(1-{Z_2 \over 6 \beta}\right)^2 \right]\ .
\label{lambdaZ}
\ee
This is equivalent to $Z_2(\lambda,\mu) = \beta Z \left( \lambda \over \beta ^3\right)$ where 
$Z\equiv Z(\lambda)$ is the generating function of the local size, which was implicitly defined as a solution of Eq.~\eqref{Z_lambda_equation}.

Considering the limit of small $\delta w$, 
we obtain
$P_{\delta w}(S,S_0) \approx \delta w \, \rho(S,S_0)$, which defines the joint density $\rho(S,S_0)$ of total and local sizes in the limit of a single avalanche. To simplify the computation, we decompose the distribution $\rho(S,S_0)$ as
\bea
\rho(S,S_0) &=& \overline{\rho}(S,S_0)+ \delta(S_0) \left( \rho(S) - \bar \rho(S) \right)\nn \\
\bar \rho(S) &=& \int_{S_0>0} \overline{\rho}(S,S_0)\ .
\label{pBarreDef}
\eea
Here $\overline{\rho}(S,S_0)$ is the smooth part of the joint density for $S$ and $S_0$, and 
is also the joint density of single avalanches containing $0$ (\textit{i.e.} $S_0 >0$). The second term takes into account all avalanches that occur away from $0$: the $\delta(S_0)$ ensures that the avalanche does not contain $0$ and the subtraction ensure that $\int_{S_0}\rho(S,S_0) = \rho(S)$ where $\rho(S)$ is   the global size density.
As we will check at the end of the calculation, the correct generating function for $\overline{\rho}$ is $Z_2(\lambda,\mu){ L^{d-1}} + 6 \left(1-\beta_\mu \right) { L^{d-1}}$.

As $\rho(S)$ is already known, we only want to compute $\overline{\rho}(S,S_0)$. 
To eliminate the term $\delta(S_0)$ we multiply (\ref{pBarreDef}) by $S_0$ and
use that $S_0 \rho(S,S_0) =S_0\overline{\rho}(S,S_0)$. Multiplication 
by $S_0$ is equivalent to taking a derivative w.r.t.\ $\lambda$ in the generating function,
\begin{widetext}
\be
\begin{split}
S_0\, \overline{\rho}(S_0,S) &= \, {  L^{d-1} }  \int_{-i \infty}^{i \infty} {d\mu \over 2 \pi i}e^{-\mu S } \int_{-i \infty}^{i \infty} {d\lambda \over 2 \pi i} e^{-\lambda S_0 } \partial_{\lambda} Z_2\left(\lambda,\mu\right) \\
&= \, {  L^{d-1} }  \int_{-i \infty}^{i \infty} {d\mu \over 2 \pi i}e^{-\mu S } \int_{-i \infty}^{i \infty} {dZ \over 2 \pi i} e^{-{\beta^3 \over 72} {Z \over \beta} \left( 6 - {Z \over \beta} \right) \left( 12 - {Z \over \beta} \right) S_0 }\ .
\end{split}
\ee
Here we changed variables from $\lambda$ to $Z_2$ (and dropped the indice) using \eqref{lambdaZ}. To simplify the calculations, we introduce a new variable $x$, s.t.\ $Z=2 \times 3^{1\over3}x+6 \beta$, with $\beta$ defined in Eq.~\eqref{defBeta}, 
\be
\begin{split}
\overline{\rho}(S_0,S) &={ L^{d-1} }\times{ 2\times 3^{1 \over 3} \over S_0} \,   \int_{-i \infty}^{i \infty} {d\mu \over 2 \pi i}e^{-\mu S } \int_{-i \infty}^{i \infty} {dx \over 2 \pi i} e^{-{x^3 \over 3} S_0 + 3^{1/3} \beta^2 x S_0}\\
&= {  L^{d-1} }\times 2\times 3^{1 \over 3}  {e^{-S/4} \over S_0} \,  \int_{-i \infty}^{i \infty} {dx\over 2 \pi i}e^{- {x^3 \over 3} S_0 } {1 \over 4} \int_{-i \infty}^{i \infty} {d y\over 2 \pi i} e^{-{y S \over 4} +(-y)^{1 / 2} 3^{1 / 3} x S_0 }\\
&= {  L^{d-1} }\times 2\times 3^{1 \over 3}  {e^{-S/4} \over S_0} \,  \int_{-i \infty}^{i \infty} {dx\over 2 \pi i}e^{- {x^3 \over 3} S_0 } \int_{0}^{ \infty} {d y \over 4 \pi} e^{- {y S \over 4}} \sin \left( \sqrt{y} 3^{1 \over 3} x S_0\right)\\
&={  L^{d-1} }\times 2\times 3^{2 \over 3} {e^{-S/4} \over \sqrt{\pi} S^{3\over 2} S_0} \,  \int_{-i \infty}^{i \infty} {dx\over 2 \pi i}e^{- {x^3 \over 3} S_0 } x S_0 e^{-{(3^{1 /3} x S_0)^2 \over S}}
\end{split}
\ee
\end{widetext}
The steps of this calculations are: first a   linear change of variable $ 4 \mu-1 \rightarrow y$, such that $\beta = (-y)^{1 \over 2}$, then a deformation of the contour of integration to integrate on both sides of the branch cut $\mathbb{R}^+$. Finally, the last integration can be performed in terms of Airy functions (e.g.\ using Appendix \ref{app:airy}),
\bea
\overline{\rho}(S,S_0)&=& \frac{6 {  L^{d-1} }}{\sqrt{\pi} S^2} e^{- {S\over 4}} F\!\left( \sqrt{3} S_0/S^{3\over 4}\right) \\
F(u) &=& {1 \over u ^{2 \over 3}} e^{-{2 \over 3} u^4 } \left( u^{ 4 \over 3} \text{Ai} \left(  u^{ 8 \over 3} \right)  -  \text{Ai}'\left( u^{ 8\over 3}\right) \right)\nn \ .
\eea
The density of avalanches with global size $S$ and which contain $0$, i.e.\ with $S_0 >0$ is
\be
\begin{split}
\overline{\rho}(S)&=\int_{0}^{\infty}\!\!\!d S_0\, \overline{\rho}(S,S_0) \\
&= {  L^{d-1} }\times 2 \sqrt{{3\over \pi}} \bigg[\int_0^{\infty} \!\!du\,F(u)\bigg] { e^{- {S\over 4}} \over S^{5 \over 4}}\\
&= {   L^{d-1} }{3 \,\Gamma\left( { 1 \over 4} \right) \over 2 \pi} {e^{-{S \over 4}} \over S^{{5 \over 4}}}
\end{split}  
\ee
where
\begin{equation}
\begin{split}
{3 \,\Gamma\left( { 1 \over 4} \right) \over 2 \pi} = 2 \sqrt{{3\over \pi}}\int_0^{\infty} du F(u) 
\approx 1.7311012158\ .
\end{split}
\end{equation}
To test our solution one can check that
\be
\int_0^\infty\!\!\!ds {3 \,\Gamma\left( { 1 \over 4} \right) \over 2 \pi} {e^{-{S \over 4}} \over S^{{5 \over 4}}} (e^{\mu S}-1) = 6 \left[1- (1-4 \mu)^{1\over 4}\right]\ .
\ee
We have checked numerically several other requirements, originating
from the definitions, namely
\bea
&&\int_0^\infty\!\!\!dS\,\bar \rho(S,S_0) = \rho_0(S_0)= \frac{2 {  L^{d-1} }}{\pi S_0} K_{1/3}\!\left(2 S_0/\sqrt{3}\right)\nn  \\
&&\int_0^\infty\!\!\!dS  \int_0^\infty\!\!\!dS_0 \,S_0 \bar \rho(S,S_0) = {  L^{d-1} }\nn \\
&&\int_0^\infty\!\!\!dS \int_0^\infty\!\!\!dS_0\, S \bar \rho(S,S_0) = 6{  L^{d-1} }\nn \\
&&\int_0^\infty\!\!\!dS \int_0^\infty\!\!\!dS_0\,  \bar \rho(S,S_0) e^{\mu S} (e^{\lambda S_0}-1) = Z_2(\lambda,\mu) {  L^{d-1} }\nn\ .
\eea 


\section{Imposed local displacement}
\label{imposeddispl} 

We set for simplicity $d=1$ in this section. 
The PDF of the global size in presence of imposed
position driving is obtained from
\be 
\overline{ e^{\mu S} } = e^{ m^2 \tilde u_{x=0} \delta w}\ ,
\ee  
where $\tilde u_x$ is the solution of a slightly modified
instanton equation:
\be 
\tilde u_x'' - m^2 \delta(x) \tilde u_x + \tilde u_x^2 = - \mu 
\ee  
and we have kept explicit the local mass. This equation is
the same as the {\it massless}  
Eq.\ (\ref{instantonJointMassless}), with $\lambda = - m^2 \tilde u_{x=0}$,
a self-consistency condition. Using its solution   given in Eqs.\ (\ref{solu2massless}) and (\ref{eqz}) we   
eliminate $\lambda$ and $z$ in the system
\be
\left\{
\begin{array}{cl}
\lambda &= - m^2 \tilde u_{x=0} = - m^2 \beta^2 \left(1- \frac{3}{2} z^2\right)  \\
\dfrac{\lambda}{\beta^3} &= 3 z (1-z^2)
\end{array}\right.
\ee
with $\beta=(- 4 \mu)^{1/4}$. It is then easy to see that there is
a solution such that $m^2 \tilde u_{x=0}$ remains finite   when 
$m^2 \to \infty$, in which case $z \to \sqrt{\frac{2}{3}}$ 
and
\be
\lim_{m^2 \to \infty} m^2 \tilde u_{x=0} = - \sqrt{\frac{2}{3}} \beta^3\ .
\ee
Hence we find
\be
P_{\delta w_0}(S) = \mbox{LT}^{-1}_{-\mu \to S}\,e^{- \delta w \sqrt{\frac{2}{3}} (-4 \mu)^{3/4}}\ .
\ee
The result for the density is simpler, 
\be \label{LT1} 
S \rho(S) = - \mbox{LT}^{-1}_{-\mu \to S}\,\partial_\mu \sqrt{\frac{2}{3}} (-4 \mu)^{3/4}\ ,
\ee 
leading to
\be
\rho(S) = \frac{\sqrt{3}}{\Gamma(1/4) S^{7/4}} 
\ee 
and a new exponent $7/4$  discussed in the main text.

\section{Some elliptic integrals for the extension distribution}
\label{app:elliptic} 
\begin{figure}[t]
\Fig{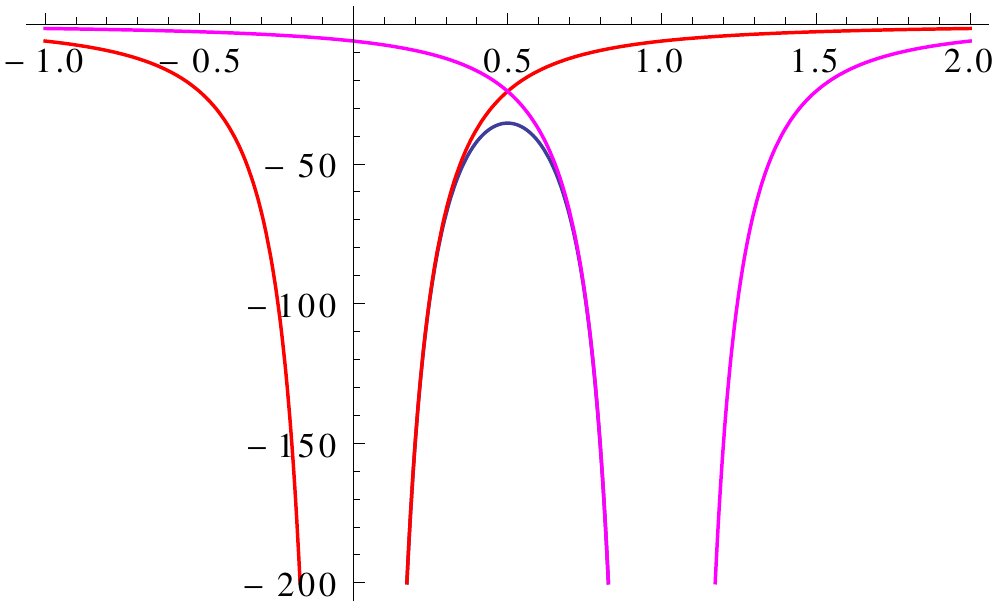}
\caption{Instanton solutions involved in the computation of $Z(r)$ for $r=1$: in blue, $\tilde{u}_1(x)$, in red $\tilde{u}_{\infty}(x)$ and in purple, $\tilde{u}_{\infty}(x-1)$.}
\end{figure}
Here we make explicit   the calculation for the density of
extensions sketched in the main text. The relevant generating function,   defined in the main text in Eq.~\eqref{GeneratingExtension}, is
\be
Z(r)=\int_x \tilde{u}_r(x)-\tilde{u}_{\infty}(x)-\tilde{u}_{\infty}(x-r)\ .
\ee
Here $\tilde{u}_r(x)$ is the solution of the instanton equation with two local sources, one at $x=0$ and one at $x=r$. The solution $\tilde{u}_{\infty}$ with one source at $x=0$ and one at infinity is equivalent to the solution with only one   source at $x=0$.

The first simplification in the calculation of this integral is the symmetry arround $r/2$. Another is that, for $x \in ]-\infty,0[$, $\tilde{u}_r(x) - \tilde{u}_{\infty}(x)$ cancels exactly. Then, the idea is to express the integral for $Z(r)$ without explicitly solving the instanton equation, using the change of variables
\be
\int \tilde u\, dx = \int \tilde u \frac{du}{\tilde u'}\ .
\label{variable_change}
\ee
This requires to express the derivative of $\tilde{u}$ w.r.t. $x$ as a function of $\tilde{u}$, which is easy because $\tilde{u}$ is solution of a differential equation, and to decompose the integral into two parts such that the change of variables is well defined: from $x= -\infty$ to $x=0$ and from $x=0$ to $x=r/2$. The rest is deduced by symmetry.

In these two intervals, $\tilde{u}_{\infty}(x-r)$ does not contain a pole, and   can  safely be computed separately. Moreover, as we said, $\tilde{u}_r(x) - \tilde{u}_{\infty}(x)$ vanishes in the first interval, i.e.\ for $x \in ]- \infty,0]$. This leaves only the integral of $\tilde{u}_r(x) - \tilde{u}_{\infty}(x)$ over $x$ running from $x=0$ to $x=r/2$. To simplify notations we introduce the variable  $t<0$, 
\be\label{deft}
t:=\tilde{u}_r(r/2)\ ,
\ee
which is in one-to-one correspondance with $r$, and is a nice parameter to express $Z$. Indeed, after the change of variables \eqref{variable_change}, the integral now runs from $u=- \infty$ to $u=t$, and  
for $0<x<r/2$, with $\tilde u \equiv \tilde{u}_r$, we have
\be
\tilde{u}_r' = \sqrt{-t^2 + \frac{2}{3}t^3 +\tilde{u}^2 - \frac{2}{3}\tilde{u}^3}\ .
\ee
Further,  with $\tilde u \equiv \tilde{u}_{\infty}$,
\be
\tilde{u}_{\infty}' = \sqrt{\tilde{u}^2 - \frac{2}{3}\tilde{u}^3}\ .
\ee
This comes from the results of Appendix~\ref{AppendixB}, and the relation $E= - t^2 + \frac{2}{3} t^3$.
To express $r$ in terms of $t$, we use the same idea as in the derivation of Eq.\ \eqref{variable_change},
\be
r = 2 \int_0^{r/2} dx = 2 \int_{-\infty}^{t} \frac{d \tilde{u}_r}{\tilde{u}_r'}\ .
\ee
Putting   these ingredients together, we obtain   $Z(r)$ as a function of $t$, which we
call $\tilde Z(t)$, in term of an elliptic integral, as well as the expression of $r$ as a function of $t$, \vfill
\begin{widetext}
\be
\begin{split}\label{F7}
\tilde Z(t)&=2 \int_{-\infty}^{t} \left( { u\over \sqrt{-t^2+{2\over3}t^3+u^2-{2\over3}u^3}} - { u \over \sqrt{u^2- {2\over 3}u^3}}\right) du - 2\int_{t}^{0}  { u \over \sqrt{u^2- {2\over 3}u^3}} du \\
&=2t \int_1^{\infty} \left( { y\over \sqrt{y^2-1-{2\over3}t(y^3-1)}} - { 1 \over \sqrt{1- {2\over 3}ty}}\right) dy -6+ 2\sqrt{9-6t}\ ,
\end{split}
\ee
\end{widetext}
\bea\label{F8}
r(t) &=& 2 \int_{-\infty}^{t} { du \over \sqrt{-t^2 +{2\over3}t^3+u^2-{2\over 3} u^3}}\nn\\
&=& 2 \int_{1}^{\infty} { dy \over \sqrt{y^2-1 -{2\over3}t(y^3-1)}}\ .
\label{def_t}
\eea
We now   use this to characterise the small-size divergence of the extension distribution. This is encoded in the small $r$ behavior of $Z(r)$, which corresponds to the large-$t$ behavior of $\tilde Z(t)$. For the latter, we have
\be
\begin{split}
Z(t) &\simeq-2 \sqrt{3\over2} \left[\int_1^{\infty}\!\!\!du \left( { u\over \sqrt{u^3-1} }- { 1 \over \sqrt{ u}}\right)  -2 \right] |t|^{1\over2}\\
& \simeq 2 \sqrt{6 \pi} {\Gamma(5/6) \over \Gamma(1/3)} |t|^{1 \over 2}\ ,
\end{split}
\ee
which is also the exact result in the massless limit. We next need to invert Eq.~\eqref{def_t} in the large-$t$ limit,
\be
 |t| \simeq  A^2 r^{-2}\; , \quad A = 2 \sqrt{6 \pi} { \Gamma(7/6) \over \Gamma(2/3) } = \sqrt{6} \frac{\Gamma(1/3)^3}{4^{2/3} \pi}\ .
\ee
The small-$r$ behavior of $Z(r)$ is then given by 
\be
Z(r) \simeq 
4 \sqrt{3 }  \pi\,r^{-1}\ .
\ee
For small $|t|$ we find
\be 
r(t) \simeq 2 \ln(12/|t|) 
\ee  
and 
\be 
\tilde Z(t) \simeq t^2 \ln(1/|t|) 
\ee  
which leads to
\be
\tilde{Z}(r) = 72 ~ r e^{-r} + O( e^{-r})
\ee
This leads to the tail of the extension density,
\be 
\rho(\ell) = \left.\partial_r^2\tilde{Z}(r)\right|_{r=\ell}  \simeq 72 \, \ell\, e^{-\ell} \text{ when } \ell \rightarrow \infty\ .
\ee  

%
%


\section{Joint distribution for extension and total size}
For simplicity, we consider only $m=0$ (massless limit). To obtain the joint distribution of extension and total size we have to add a global source $\mu$ to the instanton equation, in addition to the two local sources, whose parameters are sent to infinity. With the same tricks as previously, c.f. Appendix \ref{sec:joint} and notably Eq.~\eqref{solu1massless}, we change this problem to a new one with a mass $\beta =(-4 \mu)^{1\over4}$, but no global source. The generating function is now a function of $r$, the distance between the two local sources and $\beta$, the new mass. 
As in Appendix \ref{app:elliptic}, we can change the variable $r$ to the new parameter $t$ defined in Eq.~\eqref{deft} and express everything in terms of elliptic integrals:
\begin{widetext}
\begin{equation}
\begin{split}
r(t,\beta) &= 2\int_t^{\infty} {dy \over \sqrt{ - \beta^2t^2 - {2\over 3}t^3 + \beta^2 y^2 + {2\over 3}y^3}} = \beta^{-1} f\left( {t \over \beta^2} \right)\ ,\\
Z(t,\beta) &= - 2 \int_t^{\infty} \left( {y \over \sqrt{ - \beta^2t^2 - {2\over 3}t^3 + \beta^2 y^2 + {2\over 3}y^3}}  - {1 \over \sqrt{{2\over3}y} }\right) + 2\sqrt{ 6 t}= \beta\, g\!\left( {t \over \beta^2} \right)\ .
\end{split}
\end{equation}
The functions $f$ and $g$ are 
\begin{equation}
\begin{split}
f(x) &= 2\int_x^{\infty} {du \over \sqrt{ - x^2 - {2\over 3}x^3 + u^2 + {2\over 3}u^3}} = 2\int_1^{\infty} {du \over \sqrt{ u^2 -1 + {2\over 3}x(u^3-1)}}\ , \\
g(x) &=-2 x \int_1^{\infty} \left( {u \over \sqrt{ u^2-1 +{2\over 3}x(u^3-1)}}  - {1 \over \sqrt{{2\over3}xu} }\right) + 2\sqrt{ 6 x}\ .
\end{split}
\end{equation}
\end{widetext}
From that, we have $Z(r,\beta) = \beta \left(g \circ f^{-1} \right) (\beta r)$ and then $\partial_r^2 Z(r,\beta) = \beta^3 \left(g \circ f^{-1} \right)'' (\beta r)$ which gives
\begin{equation}
\rho(l,s)={1  \over 4 l^7} F\left( {s l^{-4} \over 4} \right)\ ,
\end{equation}
where $F$ is the inverse LT of $x\mapsto(-x)^{3 \over 4} (g\circ f^{-1})'' \left( (-x)^{1\over4} \right)$ with $g$ and $f$ the functions previously defined. Giving an analytic expression for this scaling function $F$ seems not to be possible.

\section{Numerics}
\label{a:Numerics}
We test most of our results  with a direct numerical simulation of the equation of motion \eqref{BFMdef}. This is done by discretizing both time and space. To avoid the $\sqrt{dt}$ term from a naive Euler time discretisation, we use the method of \cite{DornicChateMunoz2005},  which allows to express the exact propagator of the $d=0$ version of \eqref{BFMdef} in terms of  random distributions (Poisson and Gamma distribution). We here review  this result.

Let us start with the $d=0$ stochastic equation,
\be
\partial_t \dot{u}_t = \alpha - \beta \dot{u}_t + \sqrt{2 \sigma \dot{u}_t} \,\eta (t)
\ee
where $\eta$ is a Gaussian white noise and $\alpha$ is positive (so that $\dot{u}$ remains non-negative at all times). It can be integrated exactly using Bessel functions (\textit{cf.}\ \cite{DobrinevskiLeDoussalWiese2011b} for a derivation of this using the instanton equation for the ABBM model):
\begin{widetext}
\begin{equation}
P(\dot{u}_t | \dot{u}_0) =\frac{\beta}{\sigma} \frac{ \sqrt{\frac{\dot{u}_t}{\dot{u}_0}}^{-1+\alpha}}{{2  \sinh\left( \frac{\beta t}{2}\right)}}I_{-1+\alpha} \left( \frac{\beta}{\sigma} \frac{ \sqrt{\dot{u}_t\dot{u}_0}}{ \sinh\left( \frac{\beta t}{2}\right)}\right) \left(e^{\frac{\beta t}{2}}\right)^{\alpha}e^{-{\beta \over \sigma} \frac{\dot{u}_0 e^{-\beta t}+\dot{u}_t}{1-e^{-\beta t}}}\ .
\end{equation}
To use this representation efficiently in a numerical algorithm, the trick is to  expand it in a series, and then express it as a combination of  two distributions, 
\be
\begin{split}
P(\dot{u}_t | \dot{u}_0) =& \sum_{n=0}^{\infty} \frac{\dot{u}_t^{n-1+\alpha}\dot{u}_0^{n}}{n! \Gamma(n+\alpha)} \left( \frac{\beta}{2 \sigma \sinh \left( \frac{\beta t}{2}\right)}\right)^{2n + \alpha}\left(e^{\frac{\beta t}{2}}\right)^{\alpha} e^{-\frac{\beta}{\sigma} \frac{\dot{u}_0}{e^{\beta t}-1}} e^{-\frac{\beta}{\sigma} \frac{\dot{u}_t}{1-e^{-\beta t}}}\\
=&\sum_{n=0}^{\infty} \text{Poisson}\left[ \frac{\beta}{\sigma} \frac{ \dot{u}_0}{e^{\beta t}-1}\right](n)\, \text{Gamma}\left[n+\alpha, \frac{1-e^{-\beta t}}{\beta}\sigma \right](\dot{u}_t)\ .
\end{split}
\ee\newpage\end{widetext}

\noindent
The Poisson and Gamma distributions used above are 
\begin{align}
&\text{Poisson}\left[ \lambda \right](n) =e^{-\lambda} \frac{\lambda^n}{n!} \,\text{for } n \in \mathbb{N} \\
&\text{Gamma}\left[k, \theta \right](x) = \frac{1}{\theta (k-1)!} \left(\frac{x}{\theta}\right)^{k-1} e^{-\frac{x}{\theta}}\text{ for } x \in \mathbb{R}
\label{I4}
\end{align}
This means that we can generate $\dot{u}_t$ at time $t$ from $\dot{u}_0$ by choosing first $n$ according to the Poisson distribution and then choosing $\dot{u}_t$ from a Gamma distribution with a shape depending on $n$. This can be summed up as a nice equality between random variables,
\be 
\dot{u}_t = \text{Gamma}\left[\text{Poisson}\left[ \frac{\beta}{ \sigma} \frac{ \dot{u}_0}{e^{\beta t}-1}\right] + \alpha ,\frac{1-e^{-\beta t}}{\beta}\sigma\right] ~~~
\ee
To use this in a numerical simulation of Eq.~(\ref{BFMdef}), we first write a discretized (in space) version of the latter, 
\bea
\partial_t \dot{u}_{i,t} &=& (\dot{u}_{i+1,t} + \dot{u}_{i-1,t}) - (m^2+2) \dot{u}_{i,t} + \sqrt{2 \sigma \dot{u}_{i,t}} \xi_{i,t} \nn\\
&& +m^2 \delta w _{i,t}\ .
\eea
Choosing $ \alpha = \dot{u}_{i+1,t} + \dot{u}_{i-1,t}$, which is assumed to be constant on the time interval $[t,t+ d t]$, and $\beta = m^2+2$ in Eq.~(\ref{I4}) allows us to generate $\dot{u}_{i,t+dt}$, knowing all $\dot{u}_{i,t}$, with a correct probability distribution at order $dt$.

\section{Weierstrass and Elliptic functions}
\label{app:W} 

Here we recall  some properties of   Weierstrass's elliptic function $\mathcal{P}$
(source \cite{AbramowitzStegun} chapter 18, and Wolfram Mathworld).
It appears in complex analysis as the only doubly periodic function on the complex plane with a double 
pole $1/z^2$ at zero\footnote{It also appears as the second derivative of the Green function of the free field on a torus.}.
Denoting $\omega_1,\omega_2$ the two (a priori complex) primitive half-periods, every point of the lattice $\Lambda= \{ 2 m \omega_1 + 2 n \omega_2 | (n,m) \in \mathbb{Z}^2 \}$ is a pole of order $2$ for $\mathcal{P}$. It 
can be constructed for $z \in \mathbb{C}-\Lambda$ as
\be
\begin{split}
&{\cal P}(z|\omega_1,\omega_2):= \frac{1}{z^2} \\
& +\!\!\!\!\!\sum_{m,n \neq (0,0)} \frac{1}{(z- 2 m \omega_1 - 2 n \omega_2)^2} - \frac{1}{(2 m \omega_1+ 2 n \omega_2)^2}\ .
\end{split}
\ee
It is an even function of the complex variable $z$, with ${\cal P}(z) = {\cal P}(-z)$. Note that the choice of primitive vectors $(2 \omega_1,2 \omega_2)$ is not unique, since one can alternatively choose any linear combination. 
The conventional choice of roots $g_2$ and $g_3$ is defined from its expansion around $z=0$,
\be  \label{g2g3}
{\cal P}(z|\omega_1,\omega_2)   = \frac{1}{z^2} + \frac{g_2}{20} z^2 + \frac{g_3}{28} z^4 + \mathcal{O}(z^6)\ .
\ee  
The function $\mathcal{P}$ is alternatively denoted
\be 
 {\cal P}(z|\omega_1,\omega_2)  = {\cal P}(z;g_2,g_3) 
\ee  
the latter being defined in Mathematica as $\text{WeierstrassP}[z,\{g_2,g_3\}]$.
More explicitly, the parameters $g_2,g_3$ are expressed from the half-periods as
\bea
g_2&=& 60\!\!\!\!\! \sum_{m,n \neq (0,0)}  \frac{1}{(2 m \omega_1 + 2 n \omega_2)^4}\ ,\\
g_3 &=&  140\!\!\!\!\! \sum_{m,n \neq (0,0)}  \frac{1}{(2 m \omega_1 + 2 n \omega_2)^6}\ .
\eea 
The Weierstrass elliptic function verifies   an interesting homogeneity property,
\be \label{homo} 
\mathcal{P}(\lambda z; \lambda^{-4} g_2, \lambda^{-6} g_3) = \lambda^{-2} \mathcal{P}(z; g_2, g_3)\ ,
\ee
and the non-linear differential equation
\be \label{eqdiff1} 
{\cal P}'(z)^2 = 4 {\cal P}(z)^3 - g_2  {\cal P}(z) - g_3\ .
\ee
It is thus linked to elliptic integrals. Restricting now to $g_2,g_3 \in \mathbb{R}$ 
and focusing on $z \in \mathbb{R}$ one can   choose one half-period to be real, which we
denote   $\Omega$
\footnote{The conventions are such that if $\Delta<0$, $\Omega=\omega_1$ is real and 
$\omega_2$ imaginary (for $g_3>0$ and the 
reverse for $g_3<0$), and if $\Delta<0$, $\Omega= \omega_1 \pm \omega_2$. }.
The function ${\cal P}(z)$ is then periodic in $\mathbb{R}$ of period $2 \Omega$
and diverges at all points $2 m \Omega$, $m \in \mathbb{Z}$. It is  
defined in the fundamental interval $]0,2 \Omega[$,  repeated by 
periodicity. In this interval it satisfies the symmetry $\mathcal{P}(2 \Omega-z;g_2,g_3)=\mathcal{P}(z;g_2,g_3)$.
Its values in the first half-interval, i.e.\ for $z \in [0,\Omega]$ are such that (with $y \in [e_1,\infty]$)
\be
z=\int_y^{\infty} \frac{dt}{\sqrt{4t^3 - g_2 t - g_3}}   \Leftrightarrow y = \mathcal{P}(z;g_2,g_3)
\ee
where $e_1$ is the largest real root of the polynomial in $t$
\be
4t^3 - g_2 t - g_3 = 4 (t-e_1) (t-e_2) (t-e_3)\ .
\ee
The roots $e_i$ are all real if $\Delta=g_2^3- 27 g_3^2 >0$ and 
only one, namely $e_1$, is real if $\Delta<0$. Hence the period is given by
\be \label{int}
\Omega = \int_{e_1}^{\infty}\!\!\!\frac{dt}{\sqrt{4 t^3 - g_2 t - g_3}}\,, \; {\cal P}(\Omega)=e_1\,, \; {\cal P}'(\Omega)=0\ .
\ee
It is always finite, except when $e_1$ is a double root, in which case $\Delta=0$
and the period is infinite $\Omega=\infty$. 

For $g_2=0$ the integral (\ref{int}) can be  calculated explicitly using
\be
\begin{split}
\int_1^{\infty} \frac{du}{\sqrt{u^3-1}} &= \frac{\Gamma(1/3)^3}{4^{\frac{2}{3}} \pi}  = 
 \frac{- \sqrt{\pi}\, \Gamma(1/6)}{\Gamma(- 1/3)}\ ,\\
\int_{-1}^{\infty} \frac{du}{\sqrt{u^3+1}} &= \sqrt{\pi}\frac{\Gamma(1/3)}{\Gamma(5/6)}\ . 
\end{split}
\ee
The half-periods are
\be \label{halfper} 
\Omega=\left\{
\begin{array}{ll}
\dfrac{1}{4 \pi} \Gamma(1/3)^3 g_3^{-1/6} &\;\text{ when }\;g_3 >0\\
\sqrt{\pi} \dfrac{\Gamma(1/3)}{4^{\frac{1}{3}} \Gamma(5/6)}  |g_3|^{-1/6}  &\;\text{ when }\;g_3 <0
\end{array}\right.\ ,
\ee 
and the other period can be chosen as $\frac{1}{2} \Omega (1+ i \sqrt{3})$.

Finally, taking another derivative of (\ref{eqdiff1}) we see that the Weierstrass function also satisfies
\be  \label{second} 
{\cal P}'' (z)= 6 {\cal P}(z)^2 - \frac{g_2}{2} \ ,
\ee  
and $\mathcal{P}(z;g_2,g_3)$ is the only solution of this differential equation which satisfies (\ref{g2g3}).

From this we can find solutions of the instanton equation 
\be
\tilde u_x'' - A \tilde u_x + \tilde u_x^2 = 0\ ,
\ee
where $A=1$ is the massive case and $A=0$ the massless case. 
Comparing with Eq.~(\ref{second}) we see that a family of 
solutions are
\be  \label{soluP} 
\tilde u_x = \frac{A}{2}-6 b^2 {\cal P} \left(c+b x;\frac{A^2}{12 b^4},g_3\right)\ .
\ee  
Because of the homogeneity relation (\ref{homo}), this is a two-parameter family.
These solutions are periodic.
In the massless case $A=0$, the period of (\ref{soluP}) is $\Omega/b$ where
$\Omega$ is given by (\ref{halfper}).

\section{Non-stationary dynamics}
\label{app:nonstat} 

In the velocity theory the observables of the BFM are calculated 
from the dynamical action
\be
{\cal S}[\dot u, \tilde u] = \int_{t,q} \tilde u_{-q,t} (\partial_t + q^2 + m^2) \tilde u_{q,t} - \sigma 
\int_{t,x} \tilde u_{xt}^2 \dot u_{xt} \nn
\ee
where $\tilde u$ is the response field. The quadratic part of the action, ${\cal S}_0$, defines the free response function, 
\be
\langle \dot u_{q,t} \tilde u_{q,t'} \rangle_{{\cal S}_0} := R_{q,t-t'} =  \theta(t-t') e^{- (q^2 + m^2) (t-t')}\ .
\ee
Standard perturbation theory in the disorder $\sigma$ is then performed, and has the peculiarity
to contain only tree diagrams. It is easy to see that the average velocity 
is not corrected by the disorder, hence its value is the same as in the free theory. In presence
of a uniform driving $w=v t$,  and taking into account the initial condition $\dot u_{xt=0}=0$, 
one has
\be 
\overline{\dot u_{x,t}} = \langle \dot u_{xt} \rangle_{{\cal S}} = v \left(1- e^{- m^2 t}\right) \ .
\ee  
This implies
\begin{equation}
\overline{u_{xt}} = v t - \frac{1-e^{-m^2 t}}{m^2}\ .
\end{equation} 
Next we compute the connected correlations, where $q$   means Fourier space and $x$ real space,
\bea
\overline{\dot u_{q,t_1} \dot u_{-q,t_2}}^c &=& \langle \dot u_{q,t_1} \dot u_{-q,t_2} \rangle_{{\cal S}}\nn \\
&=& \sigma \int_{s,x} \langle \dot u_{q,t_1} \dot u_{-q,t_2} \tilde u_{x,s}^2 \dot u_{x,s}  \rangle_{{\cal S}_0} \\
 &=& 2 \sigma \int_s \langle \dot u_{xs} \rangle_{{\cal S}_0} R_{q,t_1-s} R_{q,t_2-s}\nn \ .
\eea 
Calculating this integral, and further integrating over $t_1$ and $t_2$ 
we obtain
\be 
\overline{u_{q,t} u_{-q,t}}^c = \int_0^t\!\!dt_1 \int_0^t\!\!dt_2\; \overline{\dot u_{q,t_1} \dot u_{-q,t_2}}^c \ .
\ee  
This is the final result given in the main text, see Eq.~\eqref{nonstatBFM}.

Alternatively we can obtain the correlations of $u_{xt}$ using
\bea
e^{\mu_x u_{xt_1}} && = \int_x \mu_x \overline{U_{xt_1}} + \frac{1}{2} 
\int_{x_1 x_2} \mu_{x_1} \mu_{x_2} \overline{U_{x_1 t_1} U_{x_2 t_2}}^c + ... \nn
\\ && = \exp\!\left(v m^2 \int_{x,t>0} \tilde u^\lambda_{xt}\right)
\eea 
where $\tilde u^\lambda_{xt}$ is the solution of the space-time
dependent instanton equation with a source $\lambda_{xt} = \mu_x \theta(t) \theta(t_1-t)$. 
Using the perturbation method in the source of {Section III.H} of \cite{LeDoussalWiese2012a}, specializing to that source in (261), we obtain at the end the same result as above.









\tableofcontents




\end{document}